\documentclass[twocolumn]{aa}
\usepackage{graphicx}
\usepackage{txfonts}
\def\Msun{M_\odot}
\def\Rsun{R_\odot}
\def\cmtres{\relax \ifmmode {\,\mbox{cm}}^{-3}\else \,\mbox{cm}$^{-3}$\fi}
\def\cmseis{\relax \ifmmode {\,\mbox{cm}}^{-6}\else \,\mbox{cm}$^{-6}$\fi}
\def\ergs{\relax \ifmmode {\,\mbox{erg\,s}}^{-1}\else \,\mbox{erg\,s}$^{-1}$\fi}
\def\kms{\relax \ifmmode {\,\mbox{km\,s}}^{-1}\else \,\mbox{km\,s}$^{-1}$\fi}
\def\ha{\relax \ifmmode {\mbox H}\alpha\else H$\alpha$\fi}
\def\hb{\relax \ifmmode {\mbox H}\beta\else H$\beta$\fi}
\def\hi{\relax \ifmmode {\mbox H\,{\scshape i}}\else H\,{\scshape i}\fi}
\def\hii{\relax \ifmmode {\mbox H\,{\scshape ii}}\else H\,{\scshape ii}\fi}
\def\oiii{\relax \ifmmode {\mbox O\,{\scshape iii}}\else O\,{\scshape iii}\fi}
\def\oii{\relax \ifmmode {\mbox O\,{\scshape ii}}\else O\,{\scshape ii}\fi}
\def\oi{\relax \ifmmode {\mbox O\,{\scshape i}}\else O\,{\scshape i}\fi}
\def\nii{\relax \ifmmode {\mbox N\,{\scshape ii}}\else N\,{\scshape ii}\fi}
\def\sii{\relax \ifmmode {\mbox S\,{\scshape ii}}\else S\,{\scshape ii}\fi}
\def\lha{\relax \ifmmode \mbox {L}_{H\alpha}\else $\mbox{L}_{H\alpha}$\fi}
\def\ldig{\relax \ifmmode {\mbox L}_{DIG}\else ${\mbox L}_{DIG}$\fi}
\def\ls{\relax \ifmmode {\mbox L}_{ Str}\else ${\mbox L}_{ Str}$\fi}
\def\eme{\relax \ifmmode {\,\mbox{pc\,cm}}^{-6}\else \,pc\,cm$^{-6}$\fi}
\def\l{\relax \ifmmode  \lambda\else $\lambda$\fi}

\def\etal{{et al.~}}

\def\arcmin{\hbox{$^\prime$}}
\def\arcsec{\hbox{$^{\prime\prime}$}}
\def\deg{\hbox{$^\circ$}}
\def\fs{\hbox{$^{\rm s}$}}
\def\hms#1h#2m#3s{\relax \ifmmode #1^{\rm h}\,#2^{\rm m}\,#3^{\rm s}
                   \else \hbox{$#1^{\rm h}\,#2^{\rm m}\,#3^{\rm s}$}
                  \fi}
\def\dms#1d#2m#3s{\relax#1\deg\,#2\arcmin\,#3\arcsec}
\def\hmsd#1h#2m#3.#4s{\relax\ifmmode #1^{\rm h}\,#2^{\rm m}\,#3.#4\fs
                      \else \hbox{$#1^{\rm h}\,#2^{\rm m}\,#3#4\fs$}
                      \fi}
\begin{document}
   \title{Expansive components in \hii\ regions}

%   \subtitle{I. Overviewing the $\kappa$-mechanism}

   \author{M. Rela\~no
          \inst{1}
          \and
          J. E. Beckman\inst{1,2}
%	  \fnmsep\thanks{Just to show the usage
%          of the elements in the author field}
          }

   \offprints{M. Rela\~no}

   \institute{Instituto de Astrof\'\i sica de Canarias, C. V\'\i a L\'actea s/n,
       38200, La Laguna, Tenerife, Spain \\
             \email{mpastor@ll.iac.es}
         \and
             Consejo Superior de Investigaciones Cient\'\i ficas (CSIC), Spain \\
	     \email{jeb@ll.iac.es}
%             \thanks{The university of heaven temporarily does not
%                     accept e-mails}
             }

%   \date{Received September 15, 1996; accepted March 16, 1997}

   \abstract{We study the presence of low intensity high velocity 
components, which we have termed {\it wing} features in the integrated \ha\
emission line profiles of the \hii\ region populations 
of the spiral barred galaxies NGC~1530, NGC~3359 and NGC~6951. We find 
that more than a third of the \hii\ region line profiles in each galaxy show these components. 
The highest fraction is obtained in the galaxy whose line profiles show the best S:N, 
which suggests that wing features of this type may well 
exist in most, if not all, \hii\ region line profiles.
Applying selection criteria to the wing features, 
we obtain a sample of \hii\ regions with clearly defined high velocity components in their profiles.
Deconvolution of a representative sample of the line profiles eliminates any doubt that 
the wing features could possibly be due to instrumental effects.
We present an analysis of the high velocity low intensity features
fitting them with Gaussian functions; the emission measures, central ve\-lo\-ci\-ties 
and velocity dispersions for the red and blue features take similar values. We interpret the 
features as signatures of expanding shells inside the \hii\ regions. Up to a shell 
radius of $\rm R_{shell}\sim 0.2~R_{reg}$, the stellar winds from the
central ionizing stars appear to satisfy the energy and momentum requirements
for the formation and driving the shell. 
Several examples of the most luminous \hii\ regions show that the shells appear to have larger 
radii; in these cases additional mechanisms may well be needed 
to explain the kinetic energies and momenta of the shells.

   }

   \maketitle
%
%________________________________________________________________

\section{Introduction}
The study of the internal kinematics of \hii\ regions is quite di\-ffi\-cult technically and 
has proved to be a far from easy task to achieve. The gas motions may be observed in the 
velocity profiles of the emission lines from the \hii\ regions, which show either a broad 
component, typically with supersonic width, or several components.
Previous studies have shown evidence of low intensity broad components in the 
integrated spectra of the \hii\ regions, not only in Galactic regions but also 
in those in external galaxies. Arsenault \& Roy (1986) found that a 
significant fraction of their extragalactic \hii\ region sample showed 
integrated line profiles which are better characterized by a Voigt profile 
rather than by a single Gaussian profile; while Chu \& Kennicutt (1994) and Yang et al. (1996) 
found the same broad component for nearby extragalactic \hii\ regions. 
Up to now the interpretation of these extended components has not been stu\-died for several reasons;
the low intensity of the broad components makes it difficult to confirm their 
existence, and because of this, examples of \hii\ regions showing clearly this type of 
components have been scarce until now.

The kinematics of nearby extragalactic \hii\ regions, notably 30~Doradus in the
LMC, and NGC~604 in M33, have been carefully studied and have shown the 
existence of rapid gas motions in their interiors. Smith \& Weedman (1972) first studied 
the internal kinematics of 30~Doradus with Fabry-P\'erot observations and found 
asymmetric profiles at different positions within the region; the asymmetries 
could be matched by fitting the line profile with several Gaussian components. 
Chu \& Kennicutt (1994) studied the structural kinematics of
30~Doradus using several long--slit spectra to map the velocity
field of the region. They concluded that the velocity field is dominated by a
large number of expanding structures ranging in size and expansion
velocities, most of them confined within the central zone of radius $\rm R\sim
65$~pc. In spite of the complex kinematic 
structure, the integrated velocity profile for the central part of the region, derived by adding the
profiles for all the slits, shows a broad Gaussian high intensity component with
faint extended wings.

Yang et al. (1996) made a similar study of NGC~604 combining CCD ima\-ging with long--slit 
and Fabry-P\'erot observations. They confirmed the existence of several expanding 
shells located in the central part of the region, as in 30~Doradus. The total \ha\ 
luminosity coming from these shells is $\sim$25\% of the total luminosity of 
the \hii\ region, and the
integrated ve\-lo\-ci\-ty profile of the region shows similar faint extended wings to
the profile of 30~Doradus. Other stu\-dies (Rosa \& Solf (1984); Sabalisck et al. (1995)) 
have found the same kinematic structure of expanding shells for NGC~604. 

An important question not addressed till now is whether this phenomenon can be generalized to all \hii\ 
regions. 
Up to the present time the spatial and spectral
resolution of the observations made of \hii\ regions in galaxies outside the Local Group 
have not been sufficient to allow us to obtain an answer to
this question. It is necessary to obtain a significant sample of \hii\ region
line profiles to make general inferences. However, the similarities between 
30~Doradus and NGC~604 suggest that, in general, \hii\
regions could well have comparable structures and internal kinematics.

In order to overcome these observational limitations, we present an analysis of 
the line profiles of the \hii\
region po\-pu\-la\-tions of the barred spiral galaxies NGC~1530, NGC~3359 and NGC~6951
 obtained using Fabry-P\'erot
interferometry. We confirm the existence of peaked high velocity low intensity features in a
sig\-ni\-fi\-cant fraction of the line profiles and analyze a selected sample 
of \hii\ regions to study further the nature of these extended wings. 
In providing a statistical study of the phenomenology of the wings 
we need to pay a very careful atten\-tion to eliminate doubts about the 
nature of the wings. For selected examples, we deconvolved the line profiles to obtain 
the best fit to the input spectra and to verify the existence of the low intensity components.

We interpret the wing features as signatures of an ex\-pan\-ding shell within a given \hii\ region.
The energetics of the shell and the energy input requirements to produce it are
studied for each \hii\ region of a selected sample making use of representative physical 
parameters measured in the most luminous shells observed in the nearby \hii\ regions, 
30~Doradus and NGC~604. 
Finally, we have studied quantitatively the hypothesis that the expanding shells are produced
by the interaction of the stellar winds coming from the ionizing stars with the ISM medium.

\section{Observations: H~II regions with wing features}
\subsection{Observations}
The galaxies NGC~3359, NGC~6951 and NGC~1530 
were observed with the TAURUS-II Fabry-P\'erot interferometer at the 4.2m William Herschel 
Telescope (WHT) on La Palma. The data reduction for each
galaxy is described respectively in Rozas \etal (2000b) for NGC~3359, in Rozas \etal (2002)
for NGC~6951 and in Zurita \etal (2004) for NGC~1530. 
The observations consist of exposures at a set of equally stepped separations of the 
etalon, allowing us to scan the full wavelength range of the \ha\ emission line in the galaxy.
For the three galaxies, the observed emission wavelengths were scanned in 
55 steps with exposures times at each position of the etalon of 140s for NGC~3359 
and NGC~6951, and 150s for NGC~1530. An appropiate redshifted narrow--band \ha\ filter is used as an order--sorting filter. 
Wavelength and phase calibration were performed using observations of
a calibration lamp before and after the exposure of the data cube. 
The final calibrated data cube has three axes, $x$ and $y$, corresponding to the 
spatial coordinates in the field, and $z$, which corresponds to the wavelength 
direction. The free spectral ranges for each galaxy (17.22\AA\ 
for NGC~3359 and NGC~6951 and 18.03\AA\ for NGC~1530), and the 
wavelength interval between consecutive planes (0.34\AA\ for 
NGC~3359 and NGC~6951 and 0.41\AA\ for NGC~1530), give a 
finesse of 21.5 for NGC~3359 and NGC~6951 and 21.2 for NGC~1530. 
The corresponding effective spectral resolution for our observations is close to,
and slightly less, than the 
wavelength interval between consecutive planes, which for the
data cubes from the different galaxies takes the value 15.63\kms, 
16.66\kms\ and 18.62\kms\ for NGC~6951, NGC~3359 
and NGC~1530 respectively.

From the continuum subtracted data cube we can extract line profiles 
of each point covered by the detector, thus we can obtain line profiles with 
specified apertures for any \hii\ region in the galaxy.   
We assign an \ha\ luminosity to the \hii\ regions identified in the Fabry-P\'erot 
observations using the information from the \hii\ region catalogues of these galaxies
(see Rozas \etal (1996) for NGC~6951, Rozas \etal (2000) for NGC~3359 and Rela\~no et al. (2004) for
NGC~1530).

\begin{table*}[!t]
\centering
%\hspace{-4.0cm}
%\small
\caption[]{Parameters of the wing features of the selected line profile that
  best defines the wing components from the set of spectra taken with
  different apertures for each \hii\ region in NGC~1530. Column 1:
  \hii\ region number from the catalogue (Rela\~no et al. 2004). Column 2: Logarithmic \ha\ luminosity 
  taken from the catalogue. Column 3: Aperture of the
  selected line profile. The pixel size is 0.29\arcsec, which gives a pixel scale of 51.4~pc/pix at
a galaxy distance of 36.6~Mpc (Tully 1988).
Column 4: Mean S:N of both wing components. Column 5:
Emission measure of the red component as a fraction of the total EM of the region. Column 6: Standard 
deviation of the fractional emission measures of the red Gaussian components fitted 
to the set of line profiles with different apertures taken for each \hii\
region. Columns 7 and 8: Equivalent parameters to those defined in  
columns 5 and 6 but for the blue wing components. Column 9: Velocity separation between the central 
peak and the red wing component. Column 10: Standard deviation of the quantities in column 9 for the 
set of line profiles with different apertures extracted for each \hii\
region. Columns 11 and 12: Equivalent parameters defined respectively in
columns 9 and 10 but for the blue wing component. Notes: Region 70 is included in the sample because its red wing component has $\Delta v_{\rm r}>46.55~\kms$ and Region 73 is considered to satisfy the criterion selection that its wing must have a velocity separation
bigger than 2.5 times the spectral resolution of the data cube (46.55~\kms), 
if the estimated error for $\Delta v_{\rm r}$ is considered. } 
%\vspace{0.5cm}
\begin{tabular}{llllllllllll}
\hline
\hline
\scriptsize Region &\scriptsize  log~L$_{\scriptsize\ha}$ & \scriptsize Aperture & \scriptsize S:N  
&\scriptsize EM$_{\rm r}$ & \scriptsize Stdev. 
&\scriptsize EM$_{\rm b}$ & \scriptsize Stdev. & \scriptsize $\Delta v_{\rm r}$ &\scriptsize Stdev. 
&\scriptsize $\Delta v_{\rm b}$ &\scriptsize Stdev. \\
\scriptsize (number) &\scriptsize (\ergs)   & \scriptsize pc$\times$pc &   & \scriptsize (\%) & \scriptsize (of EM$_{\rm r}$ ) 
&\scriptsize (\%) &\scriptsize (of EM$_{\rm b}$) &\scriptsize (\kms) 
&\scriptsize (of $\Delta v_{\rm r}$) &\scriptsize (\kms) &\scriptsize (of $\Delta v_{\rm b}$) \\
\hline
\scriptsize 2 &\scriptsize 39.71 &\scriptsize  155$\times$155  &\scriptsize  8.66  &\scriptsize  7.88 &
\scriptsize  0.92 &\scriptsize  4.80 &\scriptsize 1.44 &\scriptsize 56.17 &\scriptsize  0.64  
&\scriptsize 67.35 &\scriptsize  0.68\\ 
\scriptsize 6 &\scriptsize 39.64  &\scriptsize 257$\times$257  &\scriptsize 13.72  &\scriptsize  8.21 &
\scriptsize  1.56 &\scriptsize  9.40 &\scriptsize 0.92 &\scriptsize 87.60 &\scriptsize  8.07  
&\scriptsize 78.60 &\scriptsize  4.95\\
\scriptsize 7 &\scriptsize 39.59 & \scriptsize 361$\times$361  &\scriptsize 17.70  &\scriptsize 12.86 &
\scriptsize  1.98 &\scriptsize  8.61 &\scriptsize 0.99 &\scriptsize 66.59 &\scriptsize  2.53  
&\scriptsize 75.62 &\scriptsize  2.19\\ 
\scriptsize 8 &\scriptsize 39.45 & \scriptsize 257$\times$257  &\scriptsize 10.11  &\scriptsize  7.09 &
\scriptsize  0.65 &\scriptsize  9.65 &\scriptsize 2.22 &\scriptsize 73.74 &\scriptsize  1.11  
&\scriptsize 55.71 &\scriptsize  3.27\\
\scriptsize 9 &\scriptsize 39.42 & \scriptsize 154$\times$154  &\scriptsize 13.07  &\scriptsize  6.61 &
\scriptsize  1.17 &\scriptsize  7.75 &\scriptsize 2.00 &\scriptsize 72.27 &\scriptsize  2.38  
&\scriptsize 83.79 &\scriptsize 7.68\\
\scriptsize 10 &\scriptsize 39.42 & \scriptsize 257$\times$257 &\scriptsize 15.88  &\scriptsize  8.33 &
\scriptsize  1.16 &\scriptsize 11.44 &\scriptsize 0.97 &\scriptsize 69.75 &\scriptsize  2.46 
&\scriptsize 72.68 &\scriptsize  3.49\\
\scriptsize 12 &\scriptsize 39.43 & \scriptsize 154$\times$154  &\scriptsize 11.12  &\scriptsize  8.53 &
\scriptsize  1.01 &\scriptsize  7.48 &\scriptsize 2.44 &\scriptsize 96.23 &\scriptsize  11.79  
&\scriptsize 87.59 &\scriptsize 5.91\\
\scriptsize 14 &\scriptsize 39.34 &\scriptsize 257$\times$257  &\scriptsize 12.44  &\scriptsize  7.87 &
\scriptsize  0.83 &\scriptsize  6.16 &\scriptsize 0.95 &\scriptsize 87.72 &\scriptsize  6.04  
&\scriptsize 67.56 &\scriptsize  1.49\\
\scriptsize 22 &\scriptsize 39.12 &\scriptsize 257$\times$257   &\scriptsize  5.27  &\scriptsize  6.55 &
\scriptsize  1.52 &\scriptsize 7.98 &\scriptsize 0.68 &\scriptsize 49.30 &\scriptsize  7.24 
&\scriptsize 49.46 &\scriptsize 15.25\\ 
\scriptsize 27 &\scriptsize 39.05 &\scriptsize 257$\times$257  &\scriptsize 11.59  &\scriptsize 14.74 &
\scriptsize 3.40 &\scriptsize 3.78 &\scriptsize 1.48 &\scriptsize 61.38 &\scriptsize 12.85  
&\scriptsize 77.95 &\scriptsize 4.43\\          
\scriptsize 33 &\scriptsize 38.86 &\scriptsize 154$\times$154  &\scriptsize  8.79  &\scriptsize  6.73 &
\scriptsize  2.07 &\scriptsize 24.67 &\scriptsize 9.73 &\scriptsize 54.48 &\scriptsize  7.62  
&\scriptsize 53.48 &\scriptsize 6.76\\          
\scriptsize 41 &\scriptsize 38.84 &\scriptsize 263$\times$263  &\scriptsize  6.93  &\scriptsize 8.79 &
\scriptsize  1.81 &\scriptsize  17.89 &\scriptsize  2.18 &\scriptsize  61.78 &\scriptsize   1.44  
&\scriptsize  69.38 &\scriptsize 1.80\\ 
\scriptsize 70&\scriptsize 38.50&\scriptsize 154$\times$154  &\scriptsize  4.60  &\scriptsize 5.09 &
\scriptsize  1.45 &\scriptsize  12.58 &\scriptsize  2.38 &\scriptsize  63.69 &\scriptsize   2.25  
&\scriptsize  44.67 &\scriptsize 1.22\\ 
\scriptsize 73 &\scriptsize 38.42& \scriptsize 154$\times$154  &\scriptsize 12.13  &\scriptsize 15.85 &
\scriptsize  5.61 &\scriptsize  0.00 &\scriptsize 0.00 &\scriptsize 44.96 &\scriptsize  3.12  
&\scriptsize  0.00 &\scriptsize  0.00\\
\scriptsize 81 &\scriptsize 38.36& \scriptsize 257$\times$257  &\scriptsize  4.59  &\scriptsize  9.30 &
\scriptsize  0.05 &\scriptsize 4.75 &\scriptsize 0.28 &\scriptsize  85.49 &\scriptsize 7.19  
&\scriptsize  78.05 &\scriptsize  2.88\\        
\scriptsize 84 &\scriptsize 38.39& \scriptsize 154$\times$154  &\scriptsize  7.05  &\scriptsize  0.00 &
\scriptsize  0.00 &\scriptsize  16.15 &\scriptsize  3.19 &\scriptsize   0.00 &\scriptsize   0.00  
&\scriptsize  47.84 &\scriptsize  5.40\\
\scriptsize 92 &\scriptsize 38.29&\scriptsize 257$\times$257  &\scriptsize  3.72  &\scriptsize 10.37 &
\scriptsize  1.26 &\scriptsize  7.97 &\scriptsize 0.12 &\scriptsize 79.32 &\scriptsize  6.22  
&\scriptsize 78.95 &\scriptsize  2.74\\
\scriptsize 103 &\scriptsize 38.09& \scriptsize 154$\times$154  &\scriptsize  5.07 &\scriptsize 20.71 &\scriptsize  4.56 &\scriptsize   0.00 &\scriptsize  0.00 &\scriptsize  54.75 &\scriptsize   2.12  
&\scriptsize 0.00 &\scriptsize  0.00\\          
\hline
\end{tabular}
%\begin{itemize}
%\item Region 73 is considered to satisfy the criterion selection that its wing must have a velocity separation
%bigger than 2.5 times the spectral resolution of the data cube (46.55~\kms), 
%if the estimated error for $\Delta v_{\rm r}$ is considered.
%\item Region 70 is included in the sample because its red wing component has $\Delta v_{\rm r}>46.55~\kms$.
%\end{itemize}
\label{alaspar1530}
\end{table*}

\subsection{H~II regions with wing features}
A significant fraction ($\sim$70\%) of the integrated line profiles of 
the \hii\ region population in the galaxies NGC~6951, NGC~3359 and NGC~1530 are best fitted with 
two or three Gaussian components (Rela\~no et al. 2004). In this paper 
we focus our attention on the high velocity low
intensity features detected in these \hii\ region line profiles and in
particular on those secondary gaussian components located at more than $\sim 35$~\kms\ 
from the centre of the most intense Gaussian component, which we will call {\it wing} features. 
A more detailed study of the central high intensity components 
including considerations of the possible sources of broadening of 
the central peaks is presented in Rela\~no et al. (2004).

\begin{figure*}
\centering
\includegraphics[width=7.5cm]{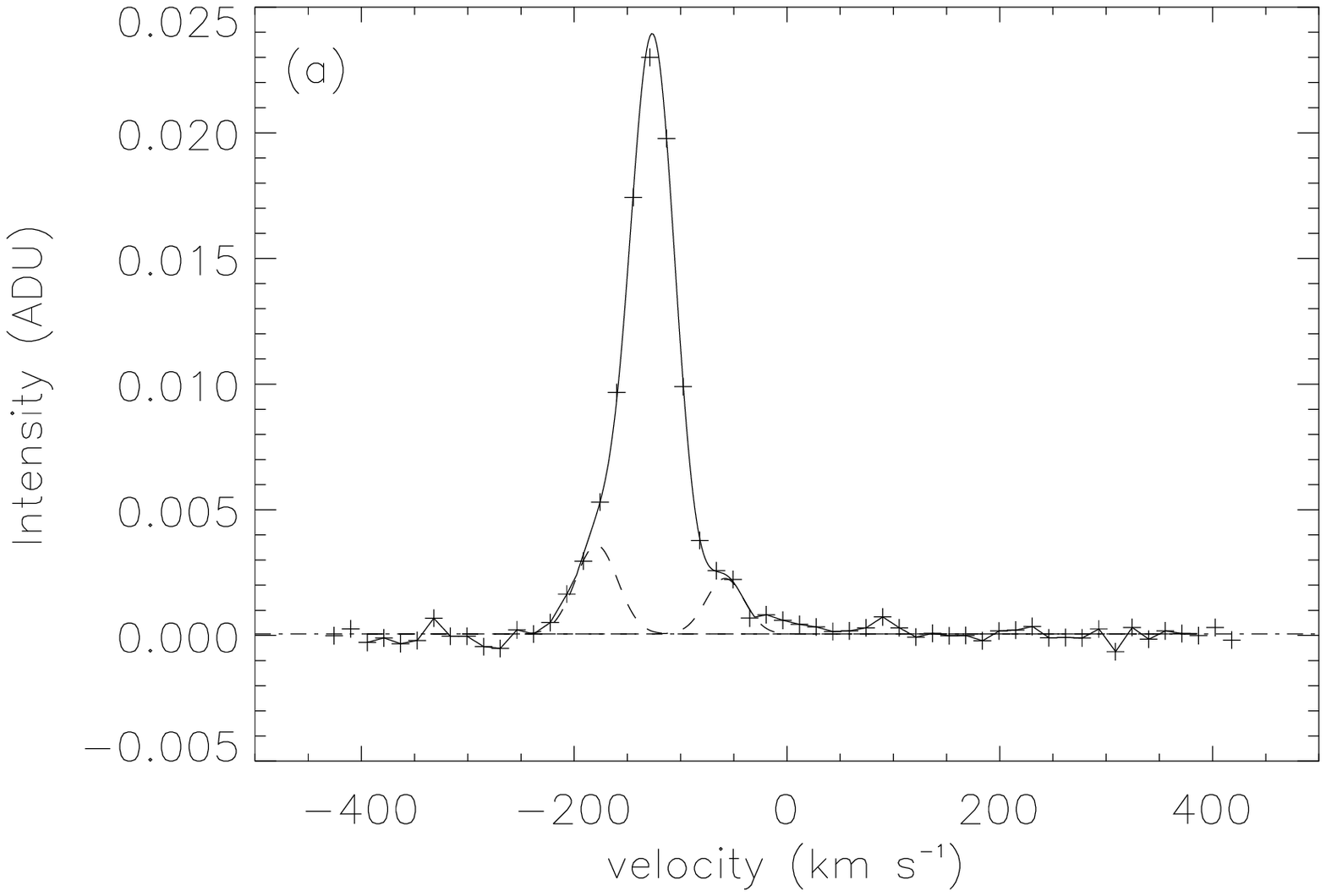}
\includegraphics[width=7.5cm]{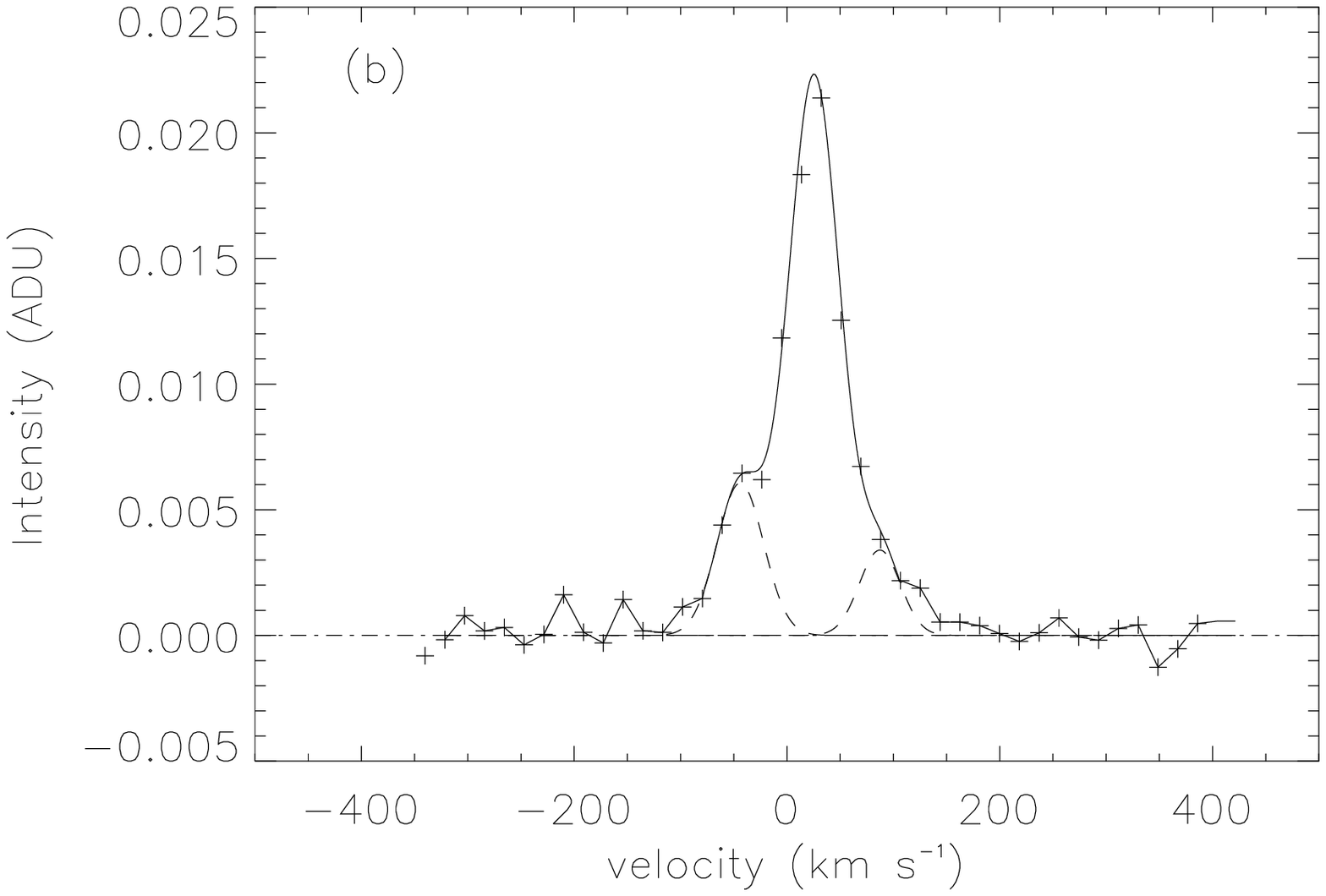}
\includegraphics[width=7.5cm]{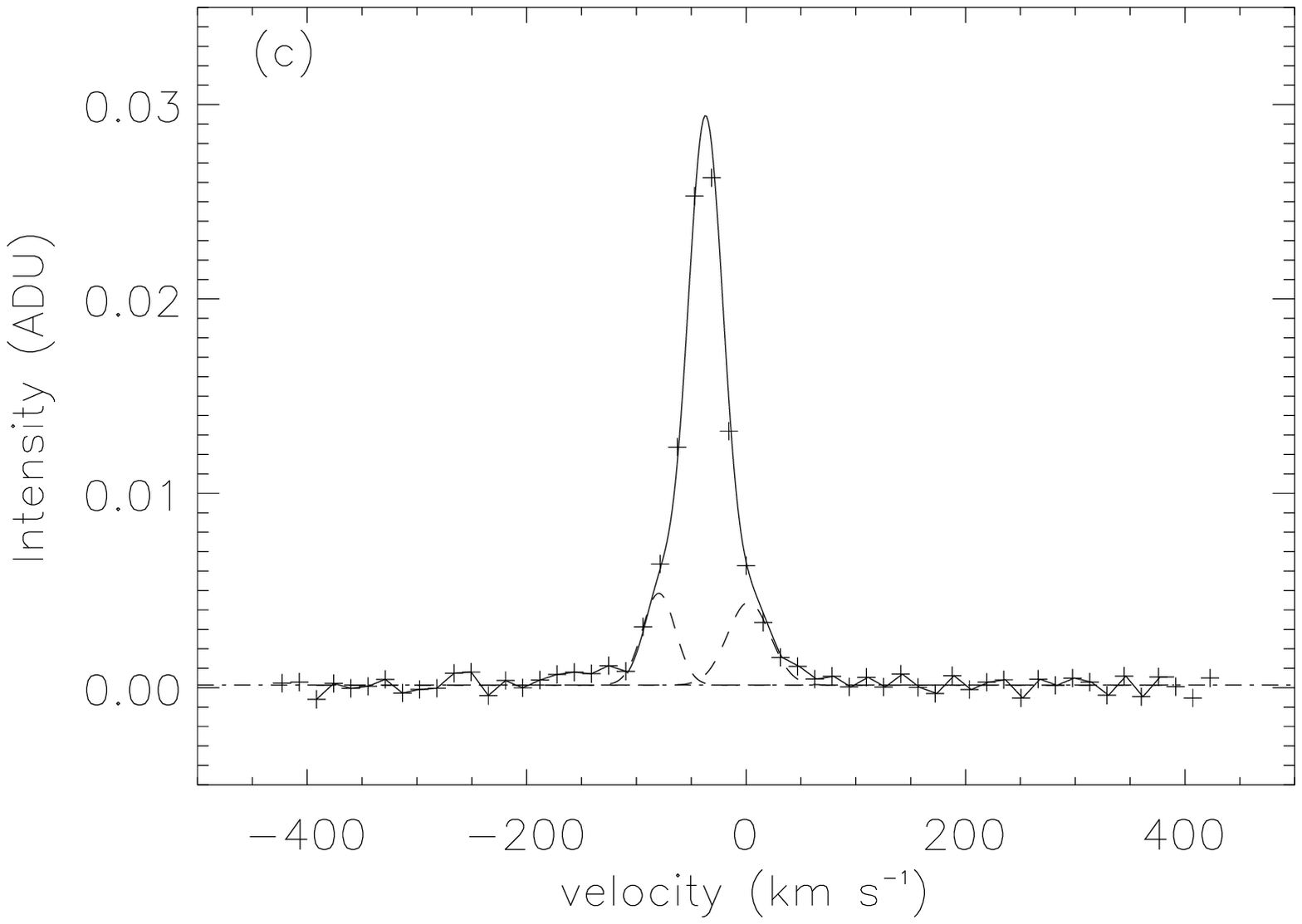}
\includegraphics[width=7.5cm]{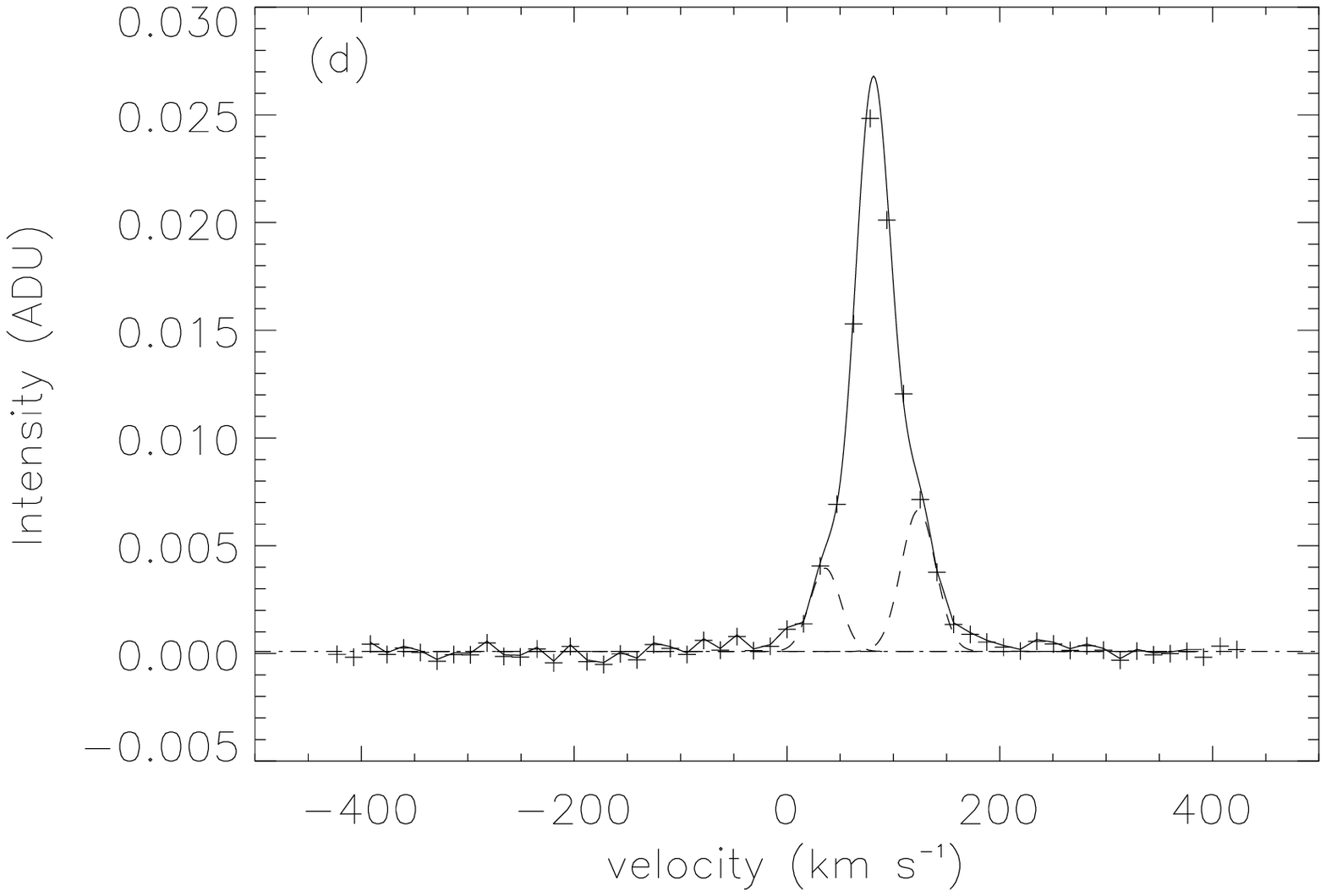}
\protect\caption[]{Examples of \hii\ region line profiles showing {\it wing} features. In dashed line we show 
the red and blue gaussian components fitted to the observed spectrum, the
continuous line show the result of the fit as a sum of the Gaussian components fitted to the spectrum.}
\label{examples_alas}
\end{figure*}

In Fig.~\ref{examples_alas} we show some examples of line profiles with this type of features. 
As we can see from this figure and from the set of line profiles analyzed, the wing features 
appear as components symmetrically shifted in velocity with respect to the central most
intense component. In order to obtain 
a sample of \hii\ region line profiles with clear detected {\it wing} features, 
we have applied the following criteria.  
{\bf a.} The amplitudes of the secondary components must be bigger than 2.5
  times the r.m.s. noise level\footnote{We define the 
root mean square noise level as the standard deviation of the points of the spectrum 
located at more than
six times the distance from the peak than the velocity dispersion of the single Gaussian 
fitted to the line profile.} of
the line profile, and {\bf b.} each secondary Gaussian component must be separated from the central 
one by more than 2.5 times the spectral 
resolution of the corres\-ponding observed data cube. We define the spectral
resolution here as the ve\-lo\-ci\-ty separation
between adjacent planes in the data cube. The values for the three galaxies are 
15.63~\kms, 16.66~\kms\ and 18.62~\kms\ for
NGC~6951, NGC~3359 and NGC~1530, respectively, which implies 
se\-pa\-ra\-tion limits in velocity of 39.08~\kms, 41.65~\kms and 46.55~\kms,
for NGC~6951, NGC~3359 and NGC~1530 respectively.  
{\bf c.} If there is a spectrum with a single secondary Gaussian component,
this must have an amplitude bigger than 3.75 times 
the root mean square noise level of the line profiles\footnote{The line profiles show a S:N$\sim$40,
thus the criterion we applied for single wings implies that their amplitudes
must be at least $\sim$10\% of the 
amplitude of the central peak.}. This last criterion 
ensures that a single wing feature can not be confused with noise. 

Down to a limit of log~L$_{\scriptsize\ha}=38.0$~(\ergs), 
the \hii\ regions whose line profiles show wing features satisfying the criteria defined above represent for 
NGC~3359, NGC~6951 and NGC~1530 respectively, 26.77\%, 42.0\% and 31.63\% of the
total number of identified \hii\ regions 
for each galaxy. The fraction of the spectra classified as 1~P+2~W or 1~P+1~W, 
(where P and W stand for central peak and wing component, res\-pec\-ti\-ve\-ly)
sastisfying the criteria are: 48.75\% for NGC~3359, 60.0\% for NGC~6951 and
81.58\% for NGC~1530. The differences bet\-ween these last three 
fractions are due to the S:N of the line profiles for each galaxy, which have
mean values at the line centres of 29.3
for NGC~3359, 41.8 for NGC~6951 and 51 for NGC~1530. The galaxy with the
highest S:N has a higher fraction of \hii\ regions with line 
profiles having secondary Gaussian components satisfying the criteria
shown above. The increase in the fraction of detectable components with S:N
ratio leads us to predict that at a sufficiently high value of S:N most, if not all,
\hii\ regions would show these wing features. 

For the sample of \hii\ regions selected, we obtained line profiles with increasing apertures; starting from a
line of sight through the centre of the region (line spectrum taken with an aperture of 
1$\times$1~pixel), we continued opening the aperture till we covered the whole
emitting area of the region and analyzed each spectrum using Gaussian
functions. We obtained a sample of isolated \hii\ regions, whose line profiles 
show wing features 
in more than one, and usually all of the line profiles of the set obtained with 
different apertures. With this procedure we guarantee that the
secondary components of the final selected sample of line
profiles show the emission from the given region, and that they are not contaminated
by the emission of neighbouring regions. 

With these considerations, we obtain a sample of 14, 13 and 19 isolated \hii\
regions for NGC~1530, NGC~6951 and NGC~3359, respectively, with clearly defined wing features
in their set of line profiles. We finally added 4 regions for NGC~1530 and NGC~6951 and 2 for NGC~3359, which were not completely
isolated but show clearly defined wings in the set of line profiles. 

\begin{table*}[!t]
\centering
\caption[]{The same parameters as in Table~\ref{alaspar1530} for the \hii\
  region line profiles of NGC~6951. For a pizel size of 0.58\arcsec\, 
the pixel scale is 67.8~pc/pix at a distance galaxy of 24.1~Mpc (Tully 1988). 
Notes: Regions 24 and 27 have a single fitted spectrum (apart from the spectrum 1$\times$1) 
of the set of the line profiles extracted with different apertures for each region. The parameters 
for the wings are those for the single fitted spectrum. Since we have only a single value for each parameter we could not obtain the corresponding standard deviation. The red wing feature of region 36 has a velocity separation less than 2.5 times the
spectral resolution of the data cube (39.08~\kms), we include it in the sample since the blue
wing does satisfy the criterion.}
%\vspace{0.5cm}
\begin{tabular}{llllllllllll}
\hline
\hline
\scriptsize Region &  \scriptsize  log~L$_{\scriptsize\ha}$     &\scriptsize Aperture & \scriptsize S:N  
&\scriptsize EM$_{\rm r}$ & \scriptsize Stdev. 
&\scriptsize EM$_{\rm b}$ & \scriptsize Stdev. & \scriptsize $\Delta v_{\rm r}$ &\scriptsize Stdev. 
&\scriptsize $\Delta v_{\rm b}$ &\scriptsize Stdev. \\
\scriptsize (number) & \scriptsize (\ergs)     & \scriptsize  (pc$\times$pc) &    & \scriptsize (\%) & \scriptsize (of EM$_{\rm r}$ ) 
&\scriptsize (\%) &\scriptsize (of EM$_{\rm b}$) &\scriptsize (\kms) 
&\scriptsize (of $\Delta v_{\rm r}$) &\scriptsize (\kms) &\scriptsize (of $\Delta v_{\rm b}$) \\
\hline
\scriptsize 2 &\scriptsize  39.27 &\scriptsize  339$\times$339  &\scriptsize  17.76 &\scriptsize  6.36 
&\scriptsize 1.14 &\scriptsize 11.09 &\scriptsize 0.29 &\scriptsize 69.88 &\scriptsize 3.91
&\scriptsize 51.98 &\scriptsize 0.69\\  
\scriptsize 4 &\scriptsize  39.11&\scriptsize  203$\times$203  &\scriptsize  10.54 &\scriptsize  8.55 
&\scriptsize  1.05 &\scriptsize  8.61  &\scriptsize  1.80 &\scriptsize 56.89 &\scriptsize  0.25
&\scriptsize  47.28 &\scriptsize 4.73\\ 
\scriptsize 6 &\scriptsize  38.97&\scriptsize  203$\times$203  &\scriptsize   8.96  &\scriptsize 10.72 
&\scriptsize  0.50 &\scriptsize  12.34 &\scriptsize  0.72 &\scriptsize 48.40 &\scriptsize  1.10 
&\scriptsize  45.79 &\scriptsize 1.34\\ 
\scriptsize 10 &\scriptsize  38.84&\scriptsize  339$\times$339   &\scriptsize  13.10  &\scriptsize  8.81 
&\scriptsize  2.20 &\scriptsize  12.76 &\scriptsize  5.27 &\scriptsize 39.10 &\scriptsize  0.59
&\scriptsize  44.35 & \scriptsize 4.41\\        
\scriptsize 11 &\scriptsize  38.82&\scriptsize  203$\times$203  &\scriptsize   10.09 &\scriptsize  6.38 
&\scriptsize 0.80 &\scriptsize 11.22 &\scriptsize 1.75 &\scriptsize 68.37 &\scriptsize 3.85
&\scriptsize 52.57 &\scriptsize 3.18\\        
\scriptsize 12 &\scriptsize  38.81 &\scriptsize  339$\times$339  &\scriptsize 4.95  &\scriptsize  10.10 
&\scriptsize  0.43 &\scriptsize 6.10 &\scriptsize  1.29 &\scriptsize 67.80 &\scriptsize  1.96
&\scriptsize  63.02 &\scriptsize 1.27\\         
\scriptsize 13 &\scriptsize  38.80 &\scriptsize  203$\times$203 &\scriptsize  10.43  &\scriptsize 11.85
&\scriptsize  1.69 &\scriptsize 6.28 &\scriptsize 0.30 &\scriptsize 43.51 &\scriptsize 1.62
&\scriptsize 50.91 &\scriptsize 1.38\\
\scriptsize 15  & \scriptsize 38.74&\scriptsize  203$\times$203 &\scriptsize   3.39  &\scriptsize  5.28 
&\scriptsize  0.55 &\scriptsize   2.58 &\scriptsize  0.29 &\scriptsize 83.23 &\scriptsize  11.13
&\scriptsize  69.49 &\scriptsize 2.23\\
\scriptsize 18  &\scriptsize 38.67 &\scriptsize  339$\times$339  &\scriptsize 7.04 & \scriptsize  3.52 
&\scriptsize  1.69 &\scriptsize  9.10 &\scriptsize  1.07 &\scriptsize 49.83 &\scriptsize  4.05
&\scriptsize  45.41 &\scriptsize  1.12\\
\scriptsize 24  &\scriptsize  38.52&\scriptsize  203$\times$203 &\scriptsize 7.59  &\scriptsize  6.92 
&\scriptsize == &\scriptsize  9.45 &\scriptsize ==&\scriptsize 54.83 &\scriptsize ==
&\scriptsize 47.06 &\scriptsize  ==\\
\scriptsize 27  &\scriptsize  38.45&\scriptsize  339$\times$339  &\scriptsize   4.73 &\scriptsize  0.00  
&\scriptsize  0.00 &\scriptsize   9.74 &\scriptsize  ==&\scriptsize   0.00 &\scriptsize  0.00
&\scriptsize  48.45 &\scriptsize  ==\\
\scriptsize 36  & \scriptsize 38.37&\scriptsize  339$\times$339 &\scriptsize 7.72  &\scriptsize  10.08 
&\scriptsize  1.29 &\scriptsize  11.39 &\scriptsize  2.69 &\scriptsize  36.40 &\scriptsize  1.75 
&\scriptsize  40.84 &\scriptsize 0.61\\         
\scriptsize 41  & \scriptsize 38.35 &\scriptsize  203$\times$203 &\scriptsize 7.08  &\scriptsize 14.28 
&\scriptsize 3.42 &\scriptsize  10.69 &\scriptsize  0.69 &\scriptsize  43.88 &\scriptsize 7.73 
&\scriptsize 57.75 &\scriptsize 4.62\\
\scriptsize 62  & \scriptsize  38.17&\scriptsize  203$\times$203 &\scriptsize  2.93  &\scriptsize  10.08 
&\scriptsize  2.05 &\scriptsize  5.52 &\scriptsize  1.44 &\scriptsize  83.50 &\scriptsize  13.04 
&\scriptsize 84.73 &\scriptsize 13.99\\         
\scriptsize 67  & \scriptsize 38.13&\scriptsize  203$\times$203 &\scriptsize   7.58  &\scriptsize  9.81 
&\scriptsize  1.31 &\scriptsize  12.24 &\scriptsize  3.94 &\scriptsize  45.21 &\scriptsize   4.09 
&\scriptsize 40.45 &\scriptsize  5.14\\         
\scriptsize 69  & \scriptsize 38.11 &\scriptsize  339$\times$339  &\scriptsize   4.13  &\scriptsize  7.95 
&\scriptsize  1.29 &\scriptsize   0.00 &\scriptsize  0.00 &\scriptsize  78.53 &\scriptsize   0.00 
&\scriptsize  0.00 &\scriptsize 0.00\\
\scriptsize 75  & \scriptsize 38.06&\scriptsize  339$\times$339 &\scriptsize   4.19  &\scriptsize 12.42 
&\scriptsize  4.19 &\scriptsize  0.00 &\scriptsize  0.00 &\scriptsize  51.53 &\scriptsize   4.28 
&\scriptsize  0.00 &\scriptsize  0.00\\
\hline
\end{tabular}
%\vspace{0.5cm}
%\begin{itemize}
%\item Regions 24 and 27 have a single fitted spectrum (apart from the spectrum 1$\times$1) 
%of the set of the line profiles extracted with different apertures for each region. The parameters 
%for the wings are those for the single fitted spectrum. Since we have only a single value for each parameter 
%we could not obtain the corresponding standard deviation.
%\item The red wing feature of region 36 has a velocity separation less than 2.5 times the
%spectral resolution of the data cube (39.08~\kms), we include it in the sample since the blue
%wing does satisfy the criterion.
%\end{itemize}
\label{alaspar6951}
\end{table*}

\subsection{Selection of the best line profile for each H~II region}
In order to obtain the parameters of the wings for each \hii\ region, we have selected from the set of 
line profiles with di\-ffe\-rent apertures that which best defines the wing features.
This is the one with the highest mean S:N for the two secondary components.
In Tables~\ref{alaspar1530},~\ref{alaspar6951} and~\ref{alaspar3359} we show
the aperture for the best line profile selected for each \hii\ region in
the three galaxies, as well as the mean S:N  
ratio, the parameters for the
red and blueshifted components, the Emission Measure (EM)\footnote{The 
value of the Emission Measure is taken as proportional to the integrated area of
the relevant fitted Gaussian component, so that the total Emission Measure 
for the \hii\ region will be proportional to the integrated
area of the complete line profile.} of the shifted
components as a fraction of the total Emission Measure of the region and the
velocity separation between the wing features and the central component. 

It is interesting to note that the parameters of the high velocity components do
not change significantly in the set of line profiles taken with different
apertures for the same \hii\ region. This can be seen from the standard
deviation values of each wing parameter shown in
Tables~\ref{alaspar1530},~\ref{alaspar6951} and~\ref{alaspar3359}, which 
are the standard deviations obtained from the values of these quantities for the spectra with 
different apertures, excluding the 1$\times$1 spectrum\footnote{We do not 
use the spectra with a 1$\times$1~pixel aperture because this aperture 
is below the limit of the angular resolution for all the three galaxies.}. 
Thus, we take the fit parameters of the best S:N line profile as the
characteristic values of the wing features for
each \hii\ region.

\begin{table*}[!t]
\centering
\hspace{-5.0cm}
%\small
\caption[]{Parameters of the wing features of the selected line profile that
  best defines the wing components from the set of spectra taken with
  different apertures for each \hii\ region in NGC~3359. The columns have the 
  same meaning as those in Table~\ref{alaspar1530}. For a pixel size 
of 0.56\arcsec\ the pixel scale is 36.4~pc/pix at a galaxy distance 
of 13.4~Mpc (de Vaucouleurs et al. 1991). Notes: Regions 140 and 192 have a 
single fitted spectrum of the set of line profiles extracted with
different apertures for each region. The parameters for the wings are 
those for the single fitted spectrum. Region 56 was kept in the sample 
although the velocity separation for its red
component is not bigger than 2.5 times the spectral resolution of the data
cube (41.65~\kms). The red components of regions 19 and 42 and the blue components 
of regions 42 and 83 satisfy the criterion if the estimated errors for 
$\Delta v_{\rm r}$ are taken into account.}
%\vspace{0.5cm}
\begin{tabular}{lllllllllllll}
\hline
\hline
\scriptsize Region & \scriptsize  log~L$_{\scriptsize\ha}$   &\scriptsize Aperture & \scriptsize S:N  
&\scriptsize EM$_{\rm r}$ & \scriptsize Stdev. 
&\scriptsize EM$_{\rm b}$ & \scriptsize Stdev. & \scriptsize $\Delta v_{\rm r}$ &\scriptsize Stdev. 
&\scriptsize $\Delta v_{\rm b}$ &\scriptsize Stdev. \\
\scriptsize (number) &  \scriptsize (\ergs)   & \scriptsize  (pc$\times$pc) &  & \scriptsize (\%) & \scriptsize (of EM$_{\rm r}$ ) 
&\scriptsize (\%) &\scriptsize (of EM$_{\rm b}$) &\scriptsize (\kms) 
&\scriptsize (of $\Delta v_{\rm r}$) &\scriptsize (\kms) &\scriptsize (of $\Delta v_{\rm b}$) \\
\hline
\scriptsize 1 &\scriptsize 39.57&\scriptsize  182$\times$182 &\scriptsize 18.61 &\scriptsize  8.06 
&\scriptsize 0.55 & \scriptsize  6.95 &\scriptsize 0.85 &\scriptsize 63.16 &\scriptsize 0.92 &\scriptsize 64.10 
&\scriptsize 6.17  \\
\scriptsize 6 & \scriptsize 39.15&\scriptsize  109$\times$109 & \scriptsize 10.78  &\scriptsize  8.11 
&\scriptsize 1.28 & \scriptsize  5.87 &\scriptsize 0.86 &\scriptsize 61.94 &\scriptsize 5.44 &\scriptsize  65.99 
&\scriptsize 2.89\\     
\scriptsize 15 &\scriptsize 38.98&\scriptsize  109$\times$109 & \scriptsize  6.07  &\scriptsize  1.69 
&\scriptsize 0.77 &\scriptsize  2.04 &\scriptsize 0.85 &\scriptsize 82.04 &\scriptsize 2.30 &\scriptsize  81.78 
& \scriptsize2.02\\     
\scriptsize 16  &\scriptsize 38.94&\scriptsize  109$\times$109 & \scriptsize  4.37  &\scriptsize  1.73 
&\scriptsize 0.40 &\scriptsize  1.55 &\scriptsize 0.48 &\scriptsize 83.87 &\scriptsize 2.86 &\scriptsize  85.25 
&\scriptsize 2.62\\     
\scriptsize 17  & \scriptsize 38.92&\scriptsize  109$\times$109 & \scriptsize  6.17  &\scriptsize  9.39 
&\scriptsize 1.52 &\scriptsize  3.04 &\scriptsize 0.60 &\scriptsize 52.48 &\scriptsize 6.00 &\scriptsize  82.12 
&\scriptsize 2.61\\     
\scriptsize 19  & \scriptsize 38.91&\scriptsize  182$\times$182  & \scriptsize 24.70  &\scriptsize 13.62 
&\scriptsize 0.98 &\scriptsize 17.44 &\scriptsize 0.67 &\scriptsize 41.50 &\scriptsize 2.39 &\scriptsize  42.20 
&\scriptsize 0.52\\       
\scriptsize 38  &\scriptsize 38.63&\scriptsize  109$\times$109 & \scriptsize 11.35  &\scriptsize  0.00 
&\scriptsize 0.00 &\scriptsize 10.55 &\scriptsize 2.73 &\scriptsize  0.00 & \scriptsize0.00 &\scriptsize  45.96 
&\scriptsize 1.44\\
\scriptsize 42  &\scriptsize 38.58&\scriptsize  109$\times$109 & \scriptsize 11.51  &\scriptsize 13.23 
&\scriptsize 4.17 &\scriptsize 11.34 &\scriptsize 2.77 &\scriptsize 41.13 &\scriptsize 2.80 &\scriptsize  40.70 
&\scriptsize 2.41\\     
\scriptsize 44  &\scriptsize 38.57&\scriptsize  109$\times$109 & \scriptsize  6.68  &\scriptsize  3.77 
&\scriptsize 3.45 &\scriptsize  2.74 &\scriptsize 0.65 &\scriptsize 68.24 &\scriptsize 3.88 &\scriptsize  50.10 
&\scriptsize 2.42\\     
\scriptsize 45  &\scriptsize 38.55&\scriptsize  109$\times$109 & \scriptsize 16.54  &\scriptsize 17.02 
&\scriptsize 3.76 &\scriptsize  8.85 &\scriptsize 1.81 &\scriptsize 42.47 &\scriptsize 1.49 &\scriptsize  45.15 
&\scriptsize 5.27\\   
\scriptsize 47  &\scriptsize 38.54&\scriptsize  109$\times$109 & \scriptsize  4.81  &\scriptsize  4.39 
&\scriptsize 0.64 &\scriptsize  3.02 &\scriptsize 0.13 &\scriptsize 70.83 &\scriptsize 1.54 &\scriptsize  83.74 
&\scriptsize 0.68 \\     
\scriptsize 51  &\scriptsize 38.52&\scriptsize  109$\times$109 & \scriptsize  9.37  &\scriptsize 10.72 
&\scriptsize 0.23 &\scriptsize 11.14 &\scriptsize 0.28 &\scriptsize 44.41 &\scriptsize 1.63 &\scriptsize  49.04 
&\scriptsize 0.94\\     
\scriptsize 52  &\scriptsize 38.51&\scriptsize  109$\times$109 & \scriptsize  9.49  &\scriptsize  7.98 
&\scriptsize 0.42 &\scriptsize  7.96 &\scriptsize 0.19 &\scriptsize 46.55 &\scriptsize 0.55 &\scriptsize  43.40 
&\scriptsize 2.13\\     
\scriptsize 53  &\scriptsize 38.49&\scriptsize  109$\times$109 & \scriptsize  4.74  &\scriptsize  4.11 
&\scriptsize 0.27 &\scriptsize  1.94 &\scriptsize 0.45 &\scriptsize 63.46 &\scriptsize 0.79 &\scriptsize  58.65 
& \scriptsize5.24\\    
\scriptsize 56  &\scriptsize 38.46&\scriptsize  109$\times$109 & \scriptsize  9.86  &\scriptsize 17.08 
&\scriptsize 2.60 &\scriptsize  5.51 &\scriptsize 0.61 &\scriptsize 37.68 &\scriptsize 1.44 &\scriptsize  51.28 
&\scriptsize 0.93\\     
\scriptsize 83  &\scriptsize 38.25&\scriptsize  109$\times$109 & \scriptsize 13.23  &\scriptsize 11.00 
&\scriptsize 0.61 &\scriptsize 15.62 &\scriptsize 0.46 &\scriptsize 41.86 &\scriptsize 0.54 &\scriptsize  40.52 
& \scriptsize1.03\\ 
\scriptsize 92  &\scriptsize 38.19&\scriptsize  109$\times$109 & \scriptsize  6.57  &\scriptsize  3.84 
&\scriptsize 0.99 &\scriptsize  7.77 &\scriptsize 0.96 &\scriptsize 58.70 &\scriptsize 1.07 &\scriptsize  45.26 
&\scriptsize 2.07\\     
\scriptsize 103 &\scriptsize 38.12&\scriptsize  109$\times$109 & \scriptsize  4.46  &\scriptsize  6.31 
&\scriptsize 0.46 &\scriptsize  0.00 &\scriptsize 0.00 &\scriptsize 55.86 & \scriptsize 1.47 &\scriptsize  0.000 
&\scriptsize 0.00\\     
\scriptsize 140 &\scriptsize 37.94&\scriptsize  109$\times$109 & \scriptsize  4.54  &\scriptsize  0.00 
&\scriptsize 0.00 & \scriptsize 18.79 &\scriptsize == &\scriptsize  0.00 &\scriptsize 0.00 &\scriptsize  51.17 
&\scriptsize ==\\ 
\scriptsize 155 &\scriptsize 37.88&\scriptsize  109$\times$109 & \scriptsize  3.42  &\scriptsize  7.34 
&\scriptsize 2.59 &\scriptsize  4.63 &\scriptsize 0.13 &\scriptsize 47.03 &\scriptsize 3.51 &\scriptsize  52.19 
&\scriptsize 0.92\\     
\scriptsize 192 &\scriptsize 37.73&\scriptsize  109$\times$109 & \scriptsize  2.63  &\scriptsize  5.30 
&\scriptsize == & \scriptsize 6.51 &\scriptsize == &\scriptsize 57.46 &\scriptsize == &\scriptsize  52.85 
&\scriptsize == \\      
\hline
\end{tabular}
%\vspace{0.5cm}
%\begin{itemize}
%\item Regions 140 and 192 have a single fitted spectrum of the set of line profiles extracted with
%different apertures for each region. The parameters for the wings are those for the single fitted spectrum. Since we
%have only a single value for each parameter we could not obtain the corresponding standard deviation.
%\item Region 56 was kept in the sample although the velocity separation for its red
%component is not bigger than 2.5 times the spectral resolution of the data
%cube (41.65~\kms). The red components of regions 
%19 and 42 and the blue components of regions 42 and 83 satisfy the criterion if the estimated errors for 
%$\Delta v_{\rm r}$ are taken into account.
%\end{itemize}
\label{alaspar3359}
\end{table*}

\subsection{Verifying the existence of the high velocity features}
\begin{figure}
\centering
\includegraphics[width=7.5cm]{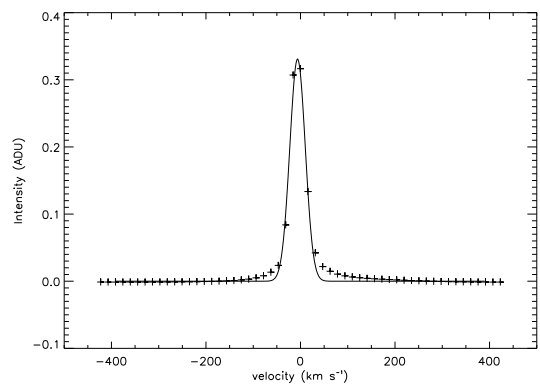}
\protect\caption[]{Normalized instrumental response of the Faby-P\'erot TAURUS-II obtained 
from the calibration lamp used in the observations of NGC~3359, shown as an example of 
Lorentzian instrument response function. The instrumental response function extracted from the 
ca\-li\-bra\-tion data cubes corresponding to the observations of NGC~6951 and NGC~1530 are similar
to that one shown here for NGC~3359.}
\label{PSF3359}
\end{figure}

The shape of the recorded spectral line is broadened by the
TAURUS--II in\-ter\-fe\-ro\-me\-ter ins\-tru\-ment function, which
shows low level extended non--Gaussian tails extending over 100~\kms\ 
from the centre of its most intense peak to the blue and the red, 
as it can be seen in Fig.~\ref{PSF3359}. These instrumental components affect the
high velocity, low intensity features detected in the spectra of the \hii\ regions.
In order to study further the physical implications of the 
high velocity components, we verified that they are
not due to the response function of TAURUS--II. To do this, we deconvolved a set
of characteristic 
observed line profiles using the method devised by Lucy (1974), which allows the deconvolution of
frequency distributions with a finite sample size.

\begin{figure*}
\centering
\includegraphics[width=7.5cm]{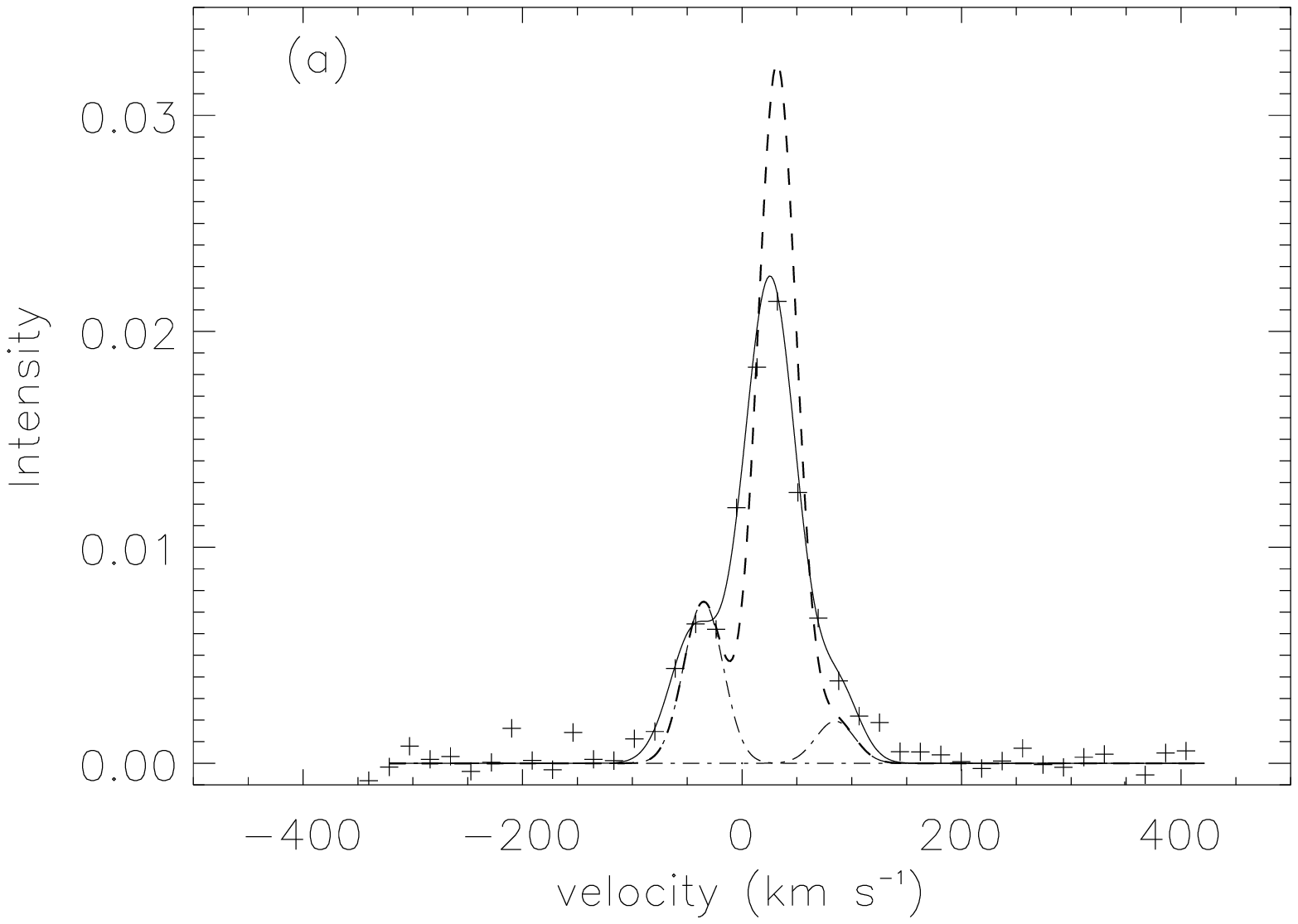}
\includegraphics[width=7.5cm]{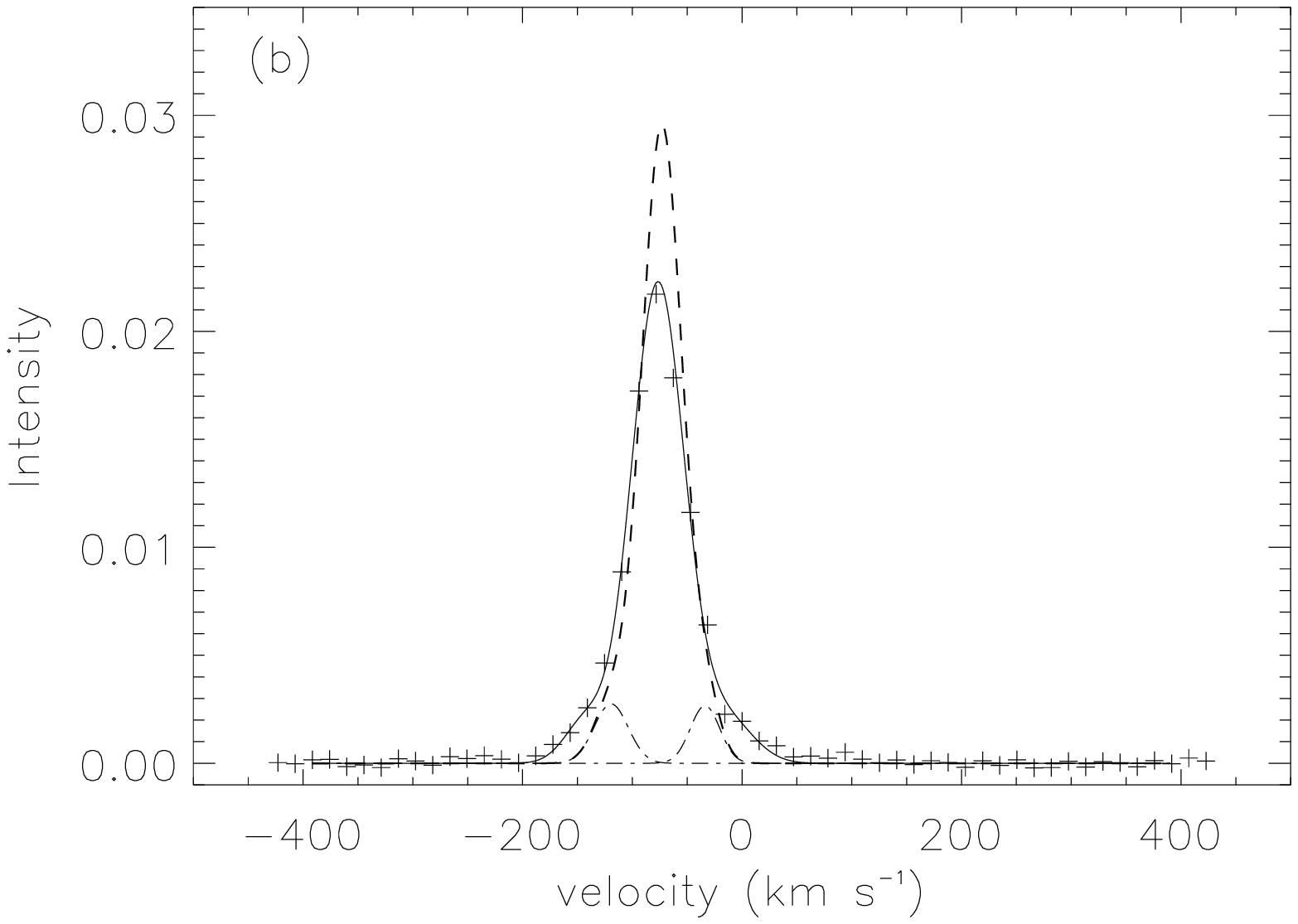}
\protect\caption[]{Deconvolution of the line profiles for (a) region 41 of
  NGC~1530 and (b) region 6 of NGC~3359. The points show the observed profile and the continous
  line represents the result of the fit to the observed data. The dashed
  line is the result of the fit of the deconvolved function obtained
  using Lucy's method and the dashed--dot line shows the low intensity
  components of the fitted deconvolved spectrum.}
\label{deconv_espectra}
\end{figure*}

\begin{table*}[t!]
\centering
%\small
\caption[]{Parameters of the Gaussian components fitted to the observed and
  deconvolved spectra of the \hii\ regions 41 of NGC~1530 and 6 of
  NGC~3359. $\rm v_{c}$, $\rm v_{r}$ and $\rm v_{b}$ are the peak
  velocities of the central, red and blue component, these velocities are
  referred to the systemic velocity of the galaxy, $\rm
  v_{sys}(NGC~1530)=(2466\pm4)~\kms$ (Zurita et al. 2003) and $\rm
  v_{sys}(NGC~3359)=(1006.8\pm0.3)~\kms$ (Rozas et al. 2000b). $\sigma_{\rm c}$,
  $\sigma_{\rm r}$ and $\sigma_{\rm b}$ are the velocity dispersions of the
  central, red and blue component, respectively.}
%\vspace{0.5cm}
\begin{tabular}{l||lll||lll||ll}
\hline
\hline
 Region 41    & $\rm v_{c}$ & $\rm v_{r}$ & $\rm v_{b}$ & $\sigma_{\rm c}$ 
             & $\sigma_{\rm r}$ & $\sigma_{\rm b}$ & EM$_{\rm r}$ & EM$_{\rm b}$  \\
\hline
 Observed     & 25.5$\pm$2.0  &
87.3$\pm$13.3 & -43.9$\pm$7.1 & 24.2$\pm$3.4 
& 19.2$\pm$13.3 & 21.9$\pm$6.6 & 8.8\% & 17.9\% \\
 Deconvolved  & 32.0$\pm$1.0  & 85.0$\pm$16.0 & -35.2$\pm$2.8 & 17.7$\pm$1.0 
& 17.8$\pm$16.7 & 18.0$\pm$3.0 & 4.7\% & 18.1\% \\
\hline
 Region 6    & $\rm v_{c}$ & $\rm v_{r}$ & $\rm v_{b}$ & $\sigma_{\rm c}$ 
             & $\sigma_{\rm r}$ & $\sigma_{\rm b}$ & EM$_{\rm r}$ & EM$_{\rm b}$  \\
\hline
 Observed     & -76.7$\pm$0.8  & -14.8$\pm$10.3 & -142.7$\pm$5.9 & 24.2$\pm$1.1 
& 24.7$\pm$6.3 & 19.1$\pm$5.1 & 8.1\% & 5.9\% \\
 Deconvolved  & -73.0$\pm$0.5  & -33.7$\pm$5.1 & -119.3$\pm$5.3 & 18.5$\pm$1.0 
& 13.4$\pm$2.7 & 15.6$\pm$3.5 & 6.9\% & 5.7\% \\
\hline
\end{tabular}
\label{Table_deconv}
\end{table*}

This method is based on
the fact that an exact reproduction of the observed function is not required
since this distribution has its own statistical fluctuations. Thus, any
solution of the deconvolved function for which the co\-rres\-pon\-ding convolved
function is close enough to the observed profile for the diffe\-ren\-ces to be
ascribable to sampling errors, is a possible solution. The ite\-ra\-ti\-ve estimation
procedure applied to this problem is that of minimizing $\chi^2$, which is a
measure of the goodness of the fit to the data, keeping normalization and
non--negativeness of the solution conserved.   

Fig.~\ref{deconv_espectra} shows the output from an
application of the Lucy algorithm to two \hii\ regions: region 41 of NGC~1530 and 
region 6 of NGC~3359. The normalized observed spectrum and the
result of its fit is shown with a continous line in these figures. The dashed line shows the
result of the fit to the deconvolved spectrum obtained using Lucy's method and 
the dashed--dot line shows the fit of the blue and redshifted components to the deconvolved function. 
The iterative procedure to obtain the deconvolved function was stopped at 
the iteration that produces a central component
in the deconvolved function with a ve\-lo\-ci\-ty dispersion close to the value
obtained by the subtraction in quadrature of the observed and the instrumental
velocity dispersions. Thus, the deconvolution of the central (and most intense)
peak is guaranteed\footnote{$\sigma_{\rm dec}(\rm
  cent)=16.8\pm0.6$~\kms for reg. 41 and $\sigma_{\rm dec}(\rm
  cent)=17.9\pm0.3$~\kms for reg. 6.}. In Table~\ref{Table_deconv} we show the fit
pa\-ra\-me\-ters for the observed and deconvolved spectra for the \hii\ regions in
Fig.~\ref{deconv_espectra}. For high velocity features which do not have
a high S:N and are close to the central most intense peak, the
result of the application of Lucy's method for deconvolution is to bring them
nearer to the centre of the main Gaussian component and to reduce their integrated
areas (see parameters of the wing components of region 6 and the red
component of region 41 in Table~\ref{Table_deconv}). When the high velocity
features have high S:N and is clearly separated from the central component, the deconvolved spectra shows the
shifted component located at a similar velocity and with similar integrated
area to the observed line profile.

The results obtained for the deconvolution of these two line
profiles show that the observed wings are not artefacts of the res\-pon\-se function of 
TAURUS--II, because the deconvolution of the observed line using the
instrument profile shows not only the central peak but also clearly reveals the
high velocity features. 

However, Lucy's method is difficult to use to extract 
the physical parameters of the shifted components because the criterion for optimizing the number of
iterations is too sensitive to the noise level of the spectrum and to the
intensity peak of the component: the S:N ratio is varying too rapidly across the
line profile to give fully consistent values for strong and weak features. For
these reasons the deconvolution with Lucy's
method cannot be used systematically where quantitative inferences are required,
but could be reliably used to verify 
the existence of the high velocity features.   

\section{Observational parameters of the wing features}
\label{Observparam}
We have studied the wing features through their fit parameters: 
area, centre and widths of the fitted Gaussian components. 
The relation between the Emission Measure, velocity and
velocity dispersion of the wing features for the best S:N line profile and the 
logarithmic \ha\ luminosity for the \hii\ regions in
Tables~\ref{alaspar1530},~\ref{alaspar6951} and~\ref{alaspar3359} are shown in
Figs.~\ref{EMvL_plot}, \ref{velvL_plot}, \ref{sigvL_plot}. There are two ways 
to determine the errors involved in the fit
parameters: first, the measured variation of the parameters in the set of line profiles taken 
with different apertures for each \hii\ region and second the estimated error of the 
parameter given by the fit program. In all the figures we show in this
section we have plotted the error bars corresponding to the first method of
error estimation mentioned above. 

We have also obtained for comparison, the mean error
of the corresponding parameter given by the fit program. 
For the three parameters: area, centre and width, the errors given by the 
fit program are bigger than the standard 
deviations of the values obtained for different apertures. For the centre 
of the Gaussian, the errors from the fit are $\sim$8~\kms, which is half of the 
spectral resolution of the observed Fabry-P\'erot data cube, while the 
standard deviation of the centre of the Gaussian fitted to the spectra with different
apertures is $\sim$3~\kms. This means two things: first,
the variation of the parameters for different apertures is small, which allows 
us to take as a canonical observational parameter that obtained 
from the fit of the best S:N line profile and, second, the S:N ratios are high
enough to allow us to resolve features with a better
resolution than the nominal spectral resolution of the data cubes, given by the separation
between planes of the data cube.
   
\subsection{Emission Measure {\it versus} log~L$_{H_\alpha}$}
\begin{figure*}
\centering
\includegraphics[width=10cm]{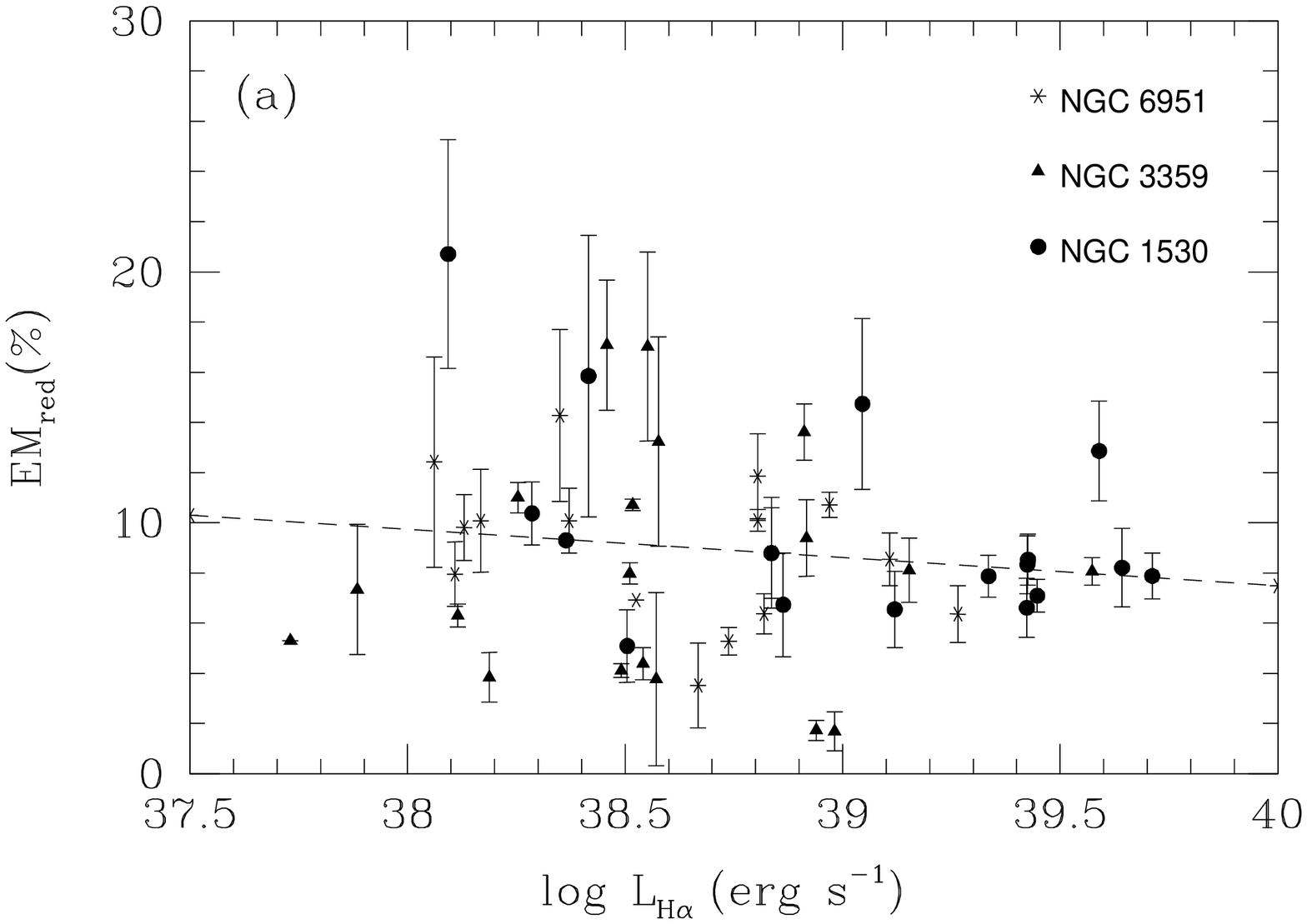}
\includegraphics[width=10cm]{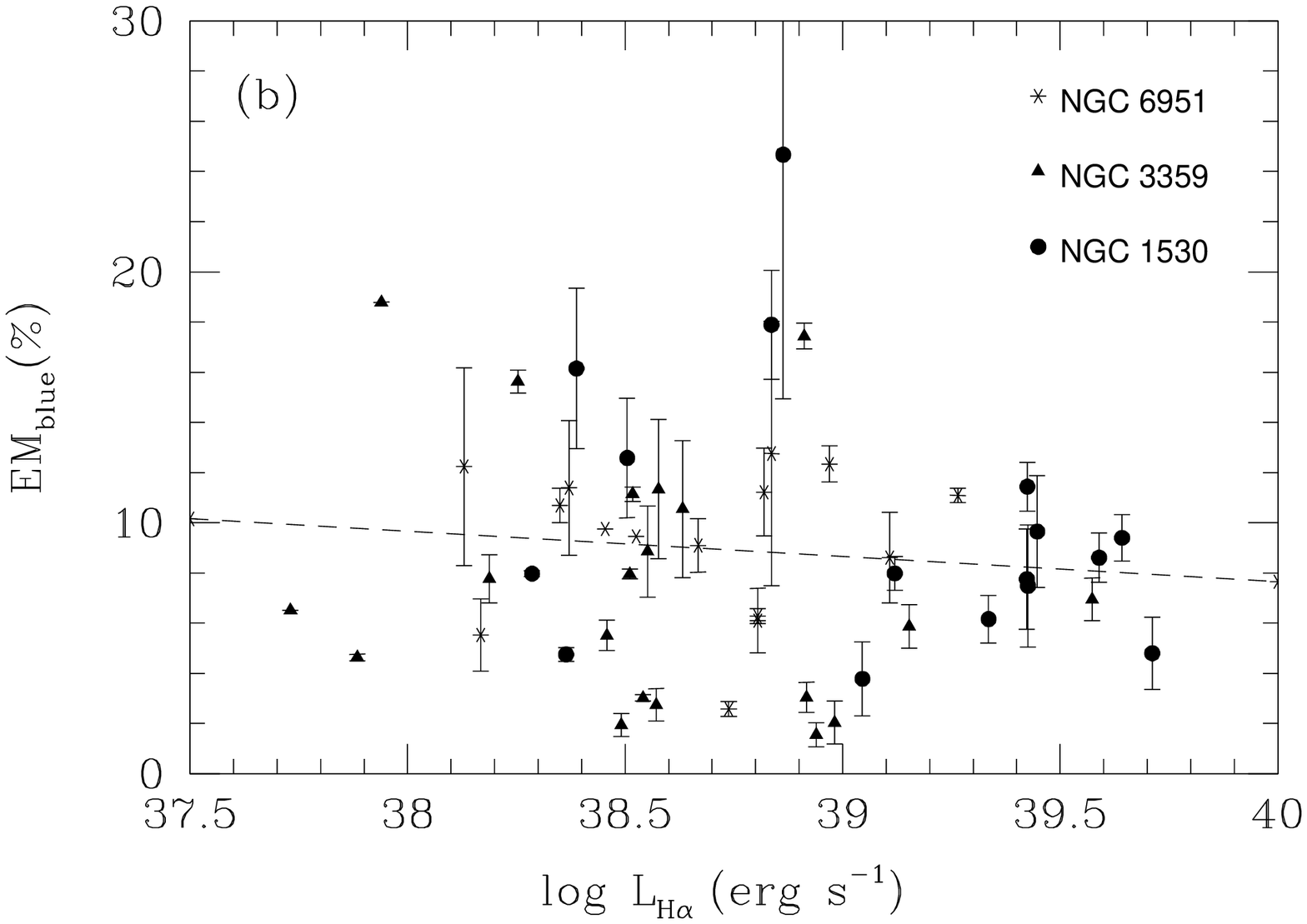}
\protect\caption[ ]{(a) Emission Measure for the red shifted Gaussian
  components fitted to the best S:N spectra of the
  selected \hii\ regions shown in
  Tables~\ref{alaspar1530},~\ref{alaspar6951} and
    \ref{alaspar3359} {\it versus.} log~\lha\ of the \hii\ region. The
    Emission Measure is given as a fraction of the total EM of the
    region. (b) Emission Measure of the blue shifted
    Gaussian component {\it versus} log~\lha\ for the same \hii\ regions 
    as in plot (a). The points without plotted error bars are for regions 
    where only a single sampling area was possible, so that no meaningful 
    measure of scatter could be derived.} 
\label{EMvL_plot}
\end{figure*}

The fractional Emission Measure of the wing features obtained as the ratio
bet\-ween the wing area and the total area of the line profile {\it versus} the 
logarithmic \ha\ luminosity is shown in Fig.~\ref{EMvL_plot}(a) for 
the red wing and Fig.~\ref{EMvL_plot}(b) for the blue wing, respectively. The
mean value for the fractional Emission
Measures of red and blue components is 8.9\% in both cases. In Figs.~\ref{EMvL_plot} 
we also show, with a dashed line the co\-rres\-pon\-ding linear fits to the data 
in the plots. The linear correlation coefficients show that there is no apparent 
relation between the fractional Emission Measures and the logarithmic 
\ha\ luminosity; $\rm r_{red}$=-0.1405 and $\rm r_{blue}$=-0.1007.

\subsection{Velocities {\it versus} log~L$_{H_\alpha}$}
\begin{figure*}
\centering
\includegraphics[width=10cm]{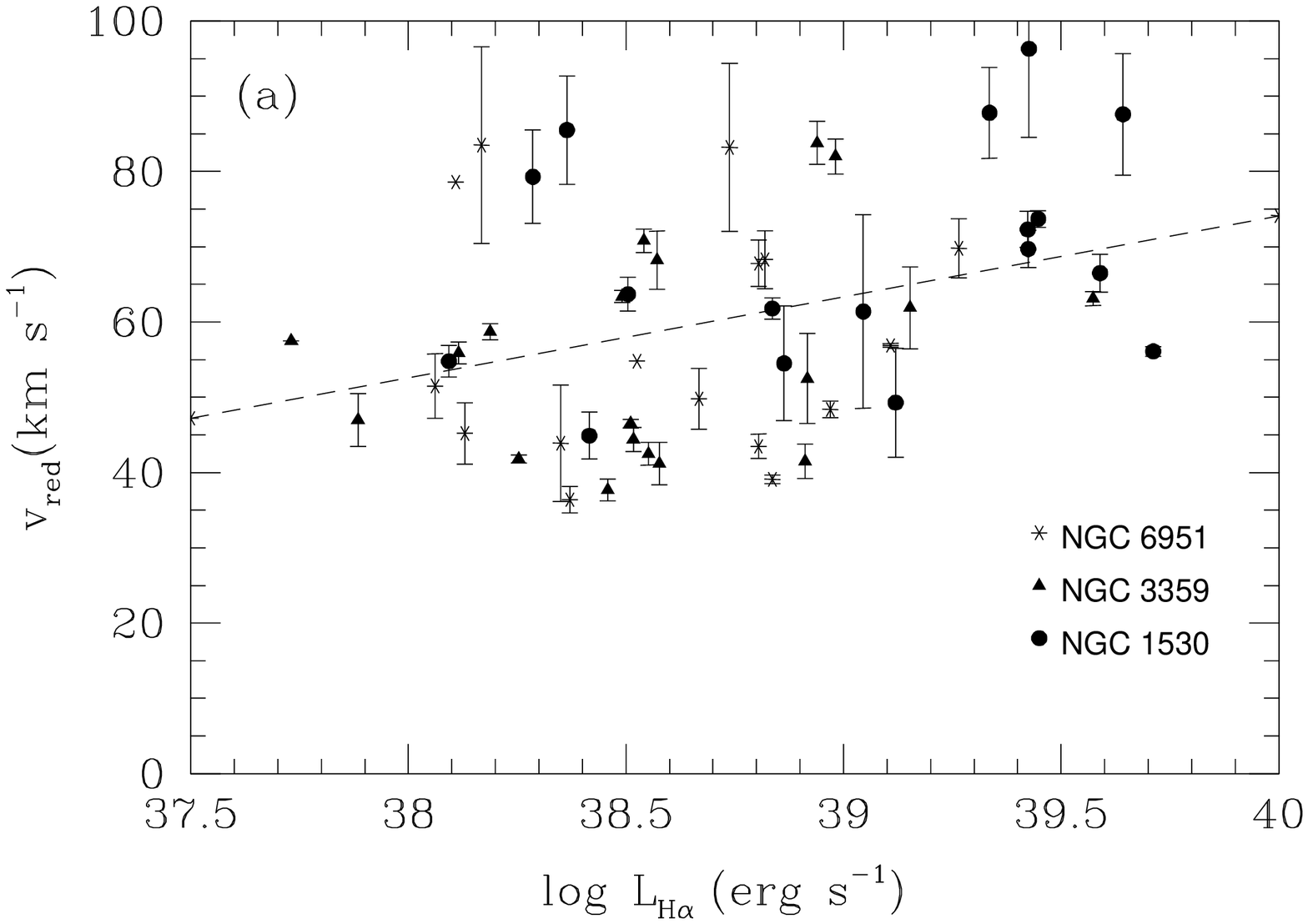}
\includegraphics[width=10cm]{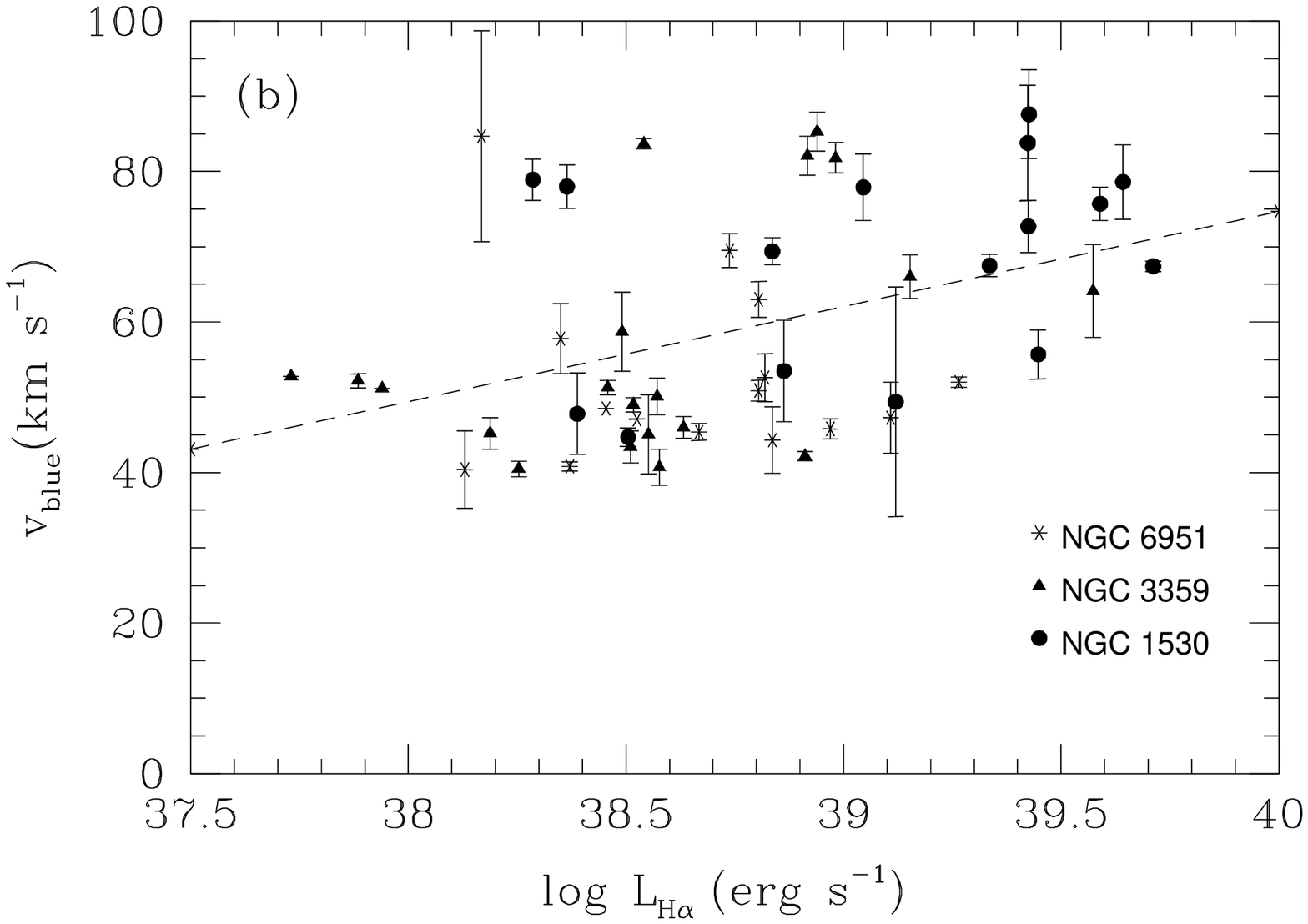}
\protect\caption[ ]{(a) Velocity separation between the red shifted component
  and the central and most intense component fitted to the best S:N spectrum
  of the selected \hii\ regions in NGC~1530, NGC~6951 and NGC~3359 {\it
    versus} log~\lha\ of the \hii\ region. (b) The same plot as in (a) but for the blue
  shifted component. For an explanation of the points with no plotted error bars see 
  Figs.~\ref{EMvL_plot}.}  
\label{velvL_plot}
\end{figure*}

In Figs.~\ref{velvL_plot} we show the velocity separation between the shifted
and most intense, i.e. the central, Gaussian components of the best S:N line profiles {\it
  versus} the logarithmic \ha\ luminosity. Figs.~\ref{velvL_plot}(a) and (b) show respectively the velocity 
  separation of the red and blue component with respect to the central peak, $\rm v_{red}$ ($\rm v_{blue}$), {\it
  versus} log~\lha. The two plots are quite similar, showing the symmetry of 
the shifted components with respect to the central and most intense component. The 
ve\-lo\-ci\-ty se\-pa\-ra\-tions range from 40~\kms\ to 90~\kms\ for these luminous \hii\ regions.
The mean value for the velocity separation is $<\rm v_{red}>=60.5$~\kms\ and 
$<\rm v_{blue}>=59.02$~\kms. There is a slight trend of increasing velocity
separation with in\-crea\-sing log~\lha, which can be seen in the linear
fits plotted in dashed lines for each parameter with linear correlation
coeffi\-cients $\rm r_{red}$=0.3417 and $\rm r_{blue}$=0.4038. 

\subsection{Wing feature velocity dispersions {\it versus} log~L$_{H_\alpha}$}
\begin{figure*}
\centering
\includegraphics[width=10cm]{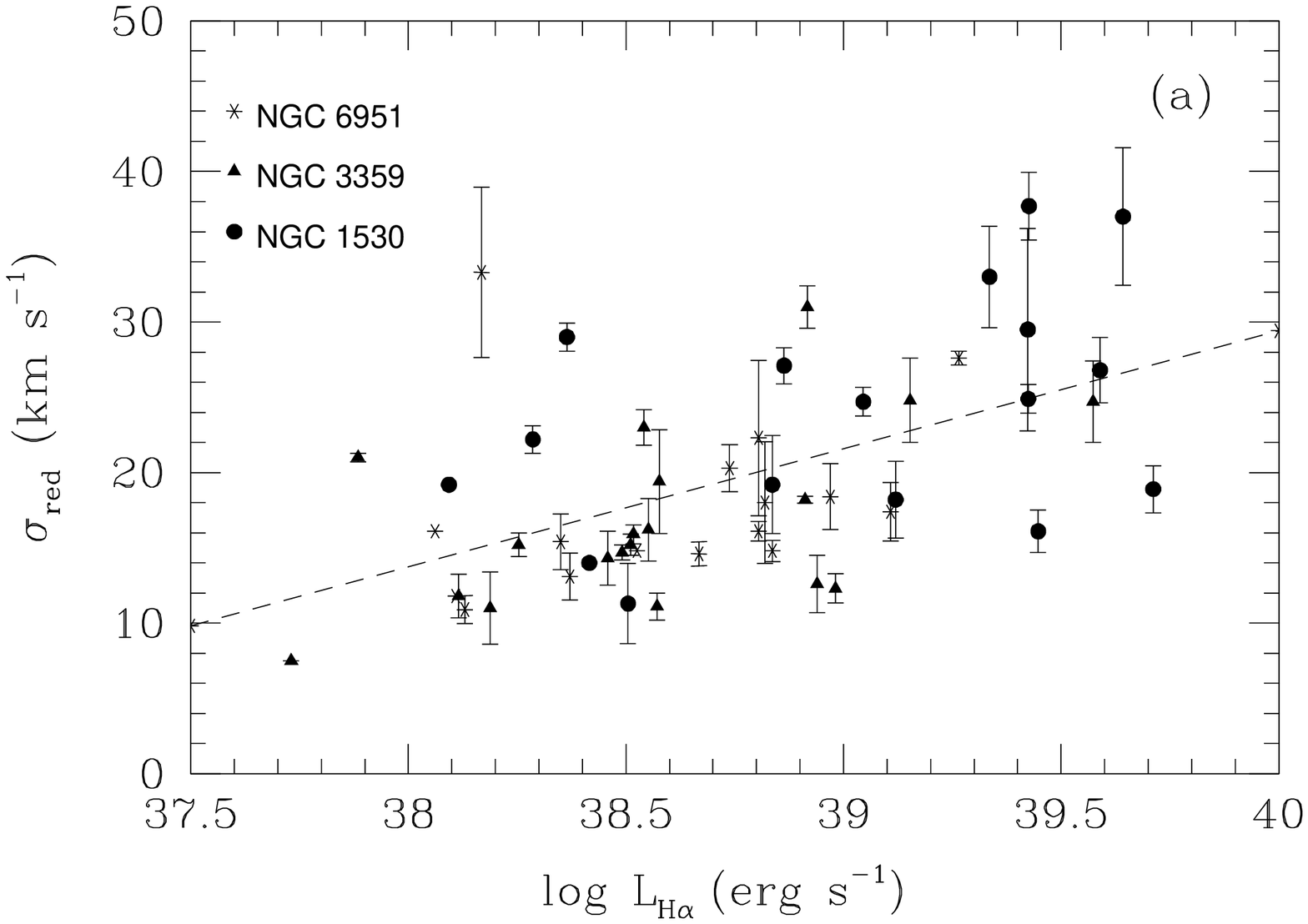}
\includegraphics[width=10cm]{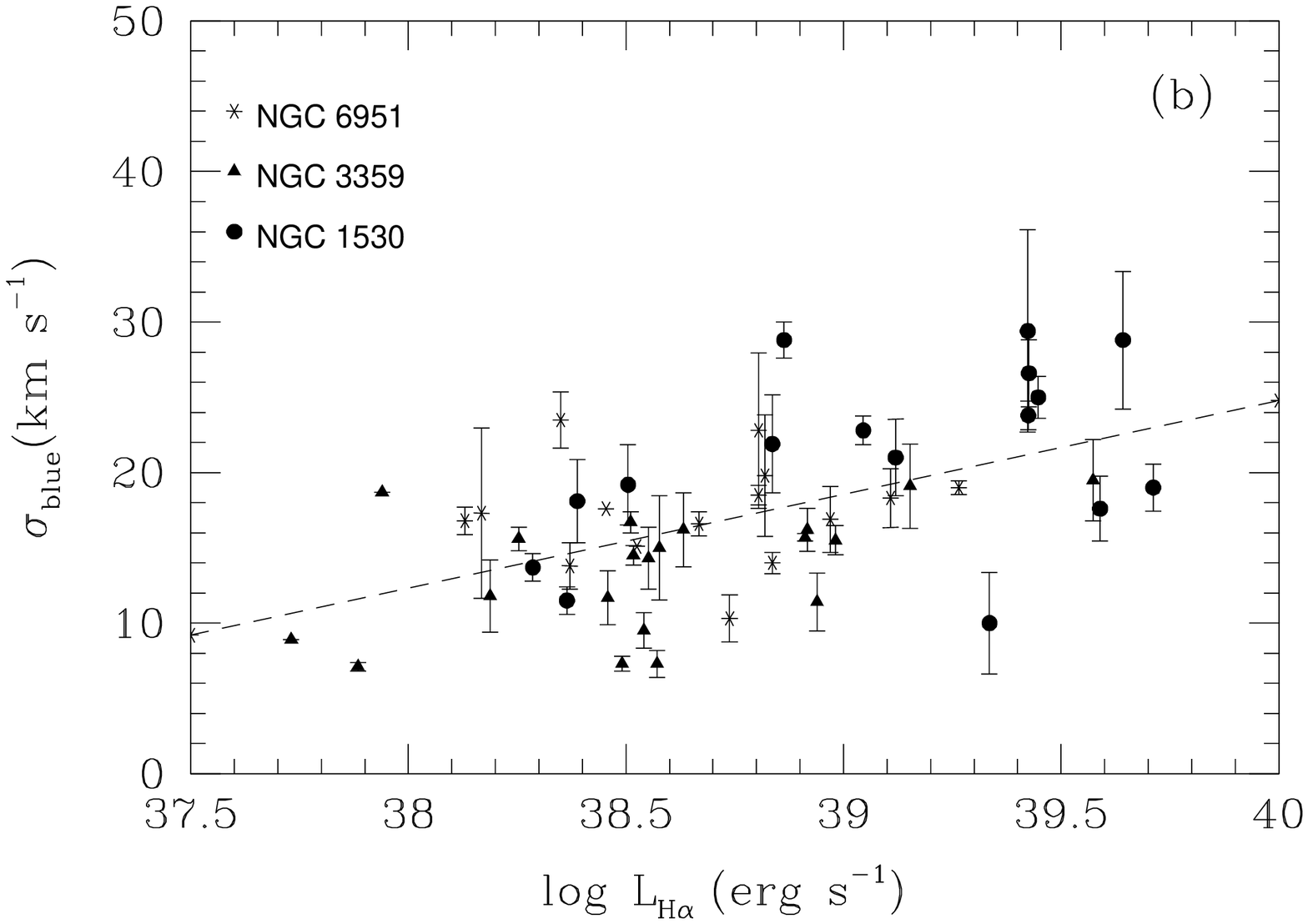}
\protect\caption[ ]{Observed velocity dispersion for the 
red (a) and blue (b) shifted component fitted to the best S:N spectrum
  of the selected \hii\ regions in NGC~1530, NGC~6951 and NGC~3359 {\it versus.}
  log~\lha\ of the \hii\ region. For an explanation of the points with no plotted
  error bars see Figs.~\ref{EMvL_plot}.}
\label{sigvL_plot}
\end{figure*}

The relation between the velocity dispersion of the wing features for the 
selected \hii\ regions and the logarithmic \ha\ luminosity is shown in
Fig.~\ref{sigvL_plot}. The red velocity dispersion is slightly higher than the 
blue one, as can be seen from com\-pa\-ri\-son of Figs.~\ref{sigvL_plot}(a) 
and~(b) and from the mean values $<\rm\sigma_{red}>=19.5$~\kms\ and
$<\rm \sigma_{blue}>=17.0$~\kms. As can be seen from
the figures, the wing velocity dispersions are low, in some cases below 
the corresponding instrumental velocity dispersion. This means that in
ge\-ne\-ral the features are not well resolved, but the fact that the
selected features satisfy the selection criteria explained in
section 2.2 imply that they are well
detected and that, although their velocity dispersions are not reliable values, 
their velocity separations and Emission Measures can be used to extract
physical information from the features.

\subsection{Velocity dispersion of the central component {\it versus}
  log~L$_{H_\alpha}$}
\begin{figure*}
\centering
\includegraphics[width=10cm]{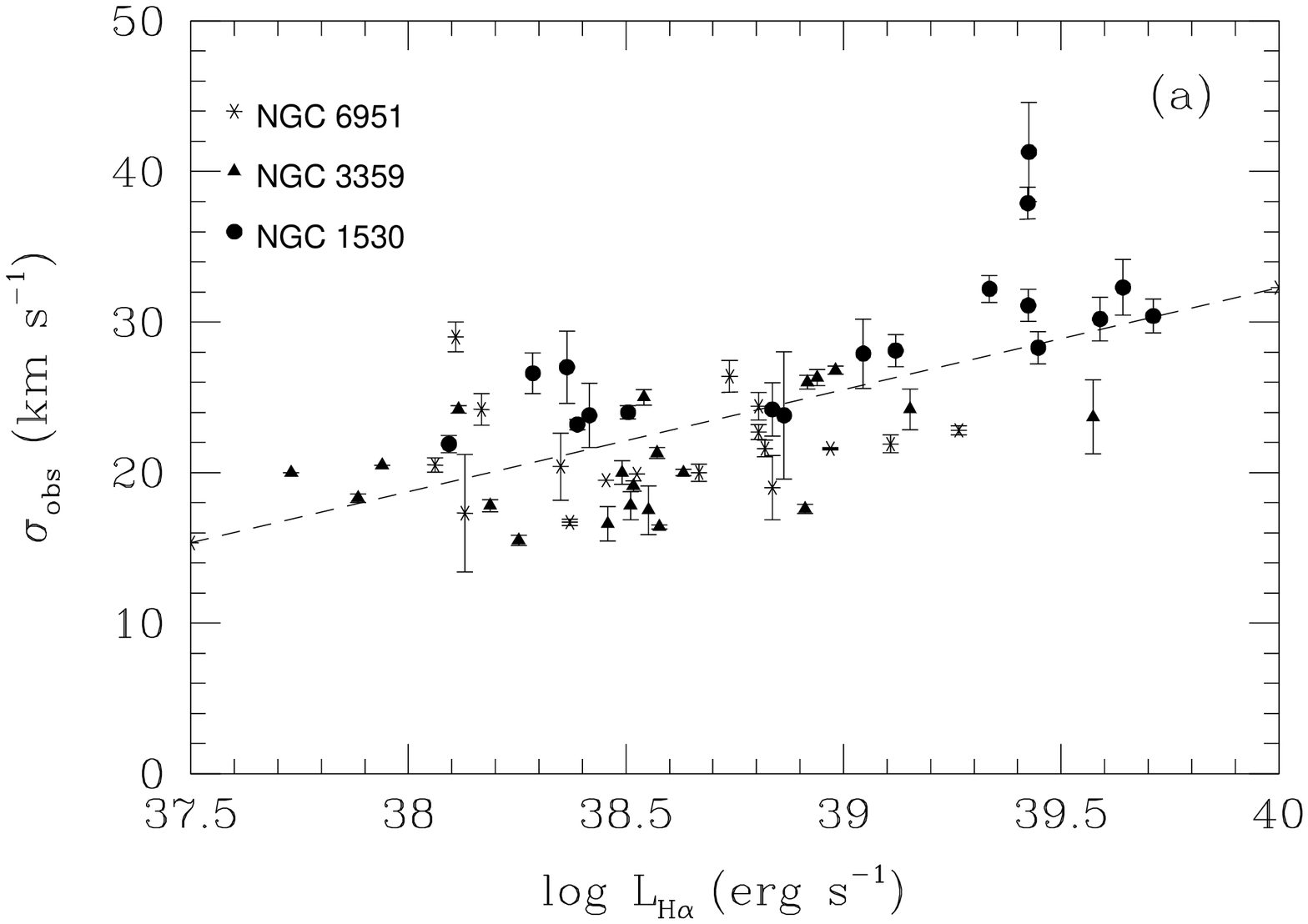}
\includegraphics[width=10cm]{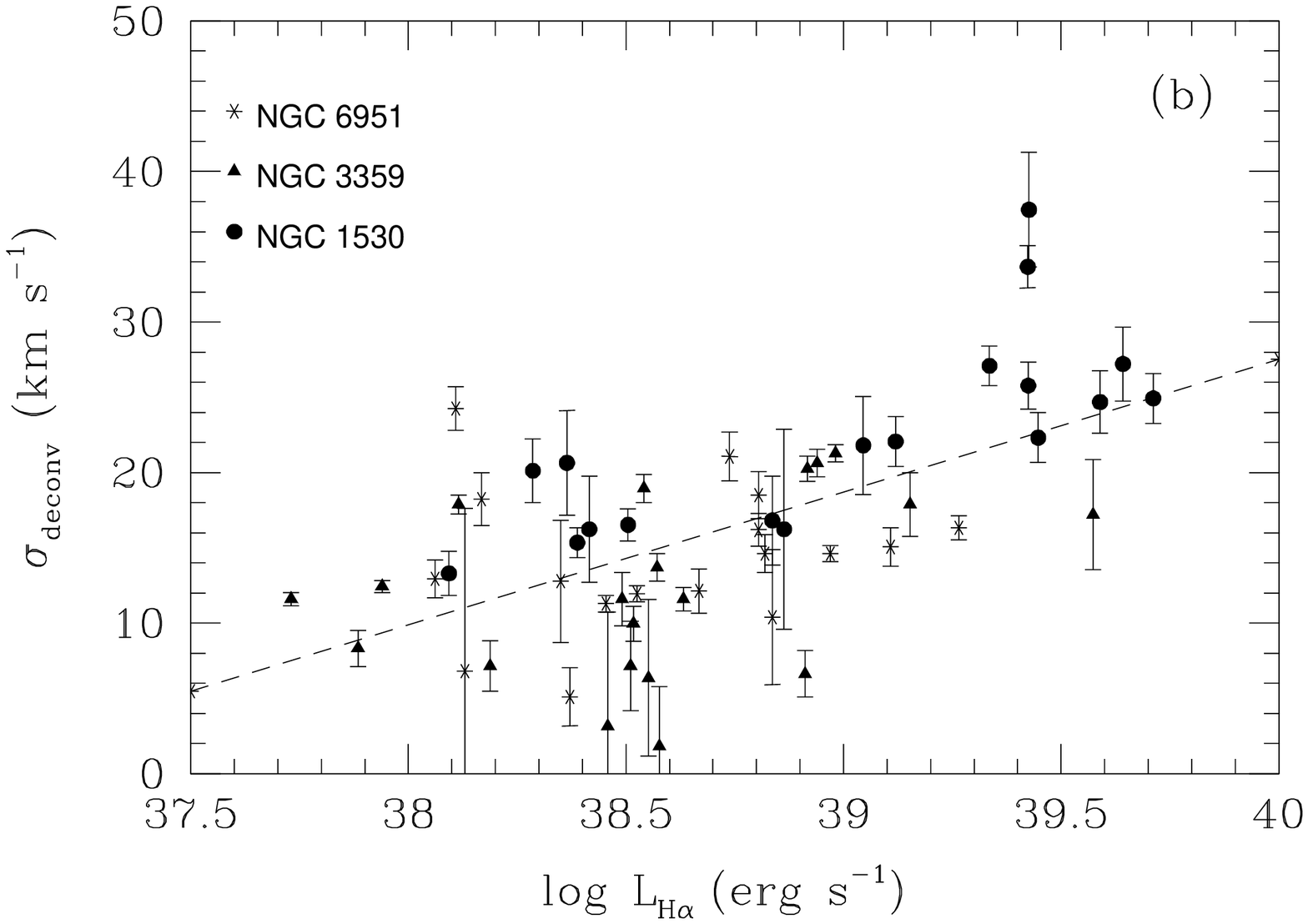}
\protect\caption[ ]{Observed (a) and deconvolved (b) velocity dispersion for the 
central Gaussian component fitted to the best S:N spectrum
  of the selected \hii\ regions in NGC~1530, NGC~6951 and NGC~3359 {\it versus.}
  log~\lha. The error bar in (a) correspond to the variation of the
  observed velocity dispersion for the set of line profiles taken for each
  \hii\ region. The error bars in (b) are estimated using the error bars in
  (a) and the corresponding errors in the instrumental velocity dispersion
  for each galaxy.}
\label{sigcentL_plot}
\end{figure*}

Fig.~\ref{sigcentL_plot}(a) shows the observed velocity dispersion for the central
Gaussian component {\it versus} the logarithmic \ha\ luminosity. There is a clear trend 
of increasing observed velocity dispersion with increasing logarithmic \ha\ luminosity, which
can be seen from the linear fit to the data points in this plot. 
The mean value of the observed velocity
dispersion for all the \hii\ regions is 23.5~\kms. 
The trend of increasing $\sigma$ with increasing log~\lha\ is also maintained 
for the {\it deconvolved}\footnote{$\rm
  \sigma_{deconv}=(\sigma^2_{obs}-\sigma^2_{inst})^{1/2}$, where
  $\rm\sigma_{inst}$ is the instrumental velocity dispersion of the data cube 
  of each galaxy.} 
velocity dispersion, as can be seen in
Fig.~\ref{sigcentL_plot}(b). However, there are some \hii\ 
regions in this plot with significantly low values of the deconvolved ve\-lo\-ci\-ty
dispersions for their central components. These
\hii\ regions show in their line profiles wing features with high
fractional Emission Measures and located relatively close in velocity to 
the central most intense component.

\section{Interpretation of the high velocity features}
We interpret the wing features observed  
as signatures of supersonic expanding shells within the \hii\ region. 
A line profile extracted for such an \hii\ region would have a central high
intensity component produced by the emission of the bulk of gas inside the
region and high velocity, low intensity features symme\-tri\-ca\-lly located with 
respect to the central peak produced by the emission of the shell. We 
use the observational parameters of these secondary components described in section~\ref{Observparam} to
des\-cri\-be the properties of the shell and to
quantify its energetics for each \hii\ region of the selected sample. 

\subsection{Shell electron density}
For a completely ionized thin shell of radius $\rm R_{shell}$ and thickness $\rm\Delta R$, 
the \ha\ luminosity of the shell will be:

\begin{equation}
\rm {L_{\scriptsize\ha}(shell)=4\pi R_{shell}^{2}\Delta R~n_{shell}^{2}
\alpha_{\scriptsize\ha}^{eff}(H,T)h\nu_{\scriptsize\ha}}
\label{equshellden}
\end{equation}
where $\rm {\alpha_{\scriptsize\ha}^{eff}(H,T)}$ is the effective recombination 
coefficient in the \ha\ emission line. From Eq.~\ref{equshellden}, 
we can obtain the shell electron density $\rm {n_{shell}}$, using the following
considerations. 
For each \hii\ region, we obtain the \ha\ luminosity of the shell multiplying the 
total \ha\ luminosity of the region given by our catalogue by the fractional area that the wings 
show in the best S:N line profile with respect to the total area of the line profile.  
We make a basic zero order assumption for the shell radius of 
$\rm {R_{shell}=0.2~R_{reg}}$, where $\rm {R_{reg}}$ is the 
\hii\ region radius, and a thickness of $\rm {\Delta R=4.5}$~pc. These values are estimates from the values 
of the most luminous expanding shells detected in 30~Doradus by Chu \& Kenicutt (1994) and 
NGC~604 by Yang et al. (1996). We cannot test the possibility that there are
shells 
with larger radii or larger thicknesses;
since it is not possible to obtain directly values of these parameters for \hii\ 
regions at the distances of the galaxies observed here. We 
have therefore used the values derived for nearby 
extragalactic \hii\ regions with similar \ha\ luminosities to those of the 
regions of our sample, which are among the most luminous regions found.

The values of the shell densities for the \hii\ region sample of NGC~1530, NGC~3359 and 
NGC~6951 are shown in co\-lumn~4 of Tables~\ref{table_energy_1530}-\ref{table_energy_6951}. 
The mean values of $\rm {n_{shell}}$ for the ga\-la\-xies are 11.5~\cmtres, 6.4~\cmtres\ and 8.2~\cmtres\ 
for NGC~1530, NGC~3359 and NGC~6951, respectively.

\begin{table*}[]
\centering
%\hspace{-3.0cm}
%\small
\caption[]{Energetics of the shell and turbulent components of the sample of \hii\ regions in NGC~1530.
Column 1: \hii\ region number from the catalogue. Column 2: Shell 
radius. Column 3: Measured shell expansion velocity. 
Column 4: Mean shell electron density. Column 5: Ionized mass of the shell. 
Column 6: Equivalent number of O3(V) type stars.  
Column 7: Kinetic energy of the shell. 
Column 8: Combined kinetic energy of OB stellar winds, assuming an
equivalent number of O3(V) type stars. Column 9: Turbulent 
kinetic energy of the body of the \hii\ region gas. Columns 10: Combined energy of the 
ionizing radiation from the region OB stars. Column 11: Combined total radiative 
energy of region OB stars emitting as black bodies.}
%\vspace{0.5cm}
%\hspace{-2.5cm}
\begin{tabular}{ccccccccccc}
\hline
\hline
\scriptsize Region & \scriptsize $\rm R_{shell}$ & \scriptsize $\rm v_{shell}$ & \scriptsize $\rm n_{shell}$ 
& \scriptsize $\rm M_{shell}$ & \scriptsize $\rm N_{eq}$ & \scriptsize $\rm E_K$ & \scriptsize $\rm E_{wind}(O3)$ 
&\scriptsize $\rm E_{turb}$ &\scriptsize  $\rm E_{rad}$ & \scriptsize $\rm Etotal_{rad}$ \\
\scriptsize (number) & \scriptsize (pc) & \scriptsize (\kms) &\scriptsize (\cmtres) & \scriptsize ($\rm 10^{4}\Msun$) 
& \scriptsize (O3(V)) &
\scriptsize ($10^{51}$~erg) & \scriptsize ($10^{51}$~erg) & \scriptsize ($10^{51}$~erg) & 
\scriptsize ($10^{54}$~erg) & \scriptsize ($10^{54}$~erg) \\
\hline
2  &  88.64 & 61.76 & 11.90 & 13.1  & 50.61 & 5.0 & 19.2 &  60.0  & 2.6  & 6.6 \\
6  &  87.66 & 83.10 & 13.08 & 14.0  & 43.10 & 9.6 & 16.3 &  62.6  & 2.2  & 5.6 \\
7  &  85.12 & 71.11 & 14.01 & 14.2  & 38.20 & 7.1 & 14.5 &  44.2  & 2.0  & 5.0 \\
8  &  82.00 & 64.72 & 10.90 & 10.2  & 27.53 & 4.3 & 10.4 &  25.0  & 1.4  & 3.6 \\
9  &  69.92 & 78.03 & 11.51 & 7.9   & 26.05 & 4.8 &  9.9 &  60.7  & 1.3  & 3.4 \\
10 &  74.16 & 71.21 & 12.76 & 9.8   & 26.12 & 4.9 &  9.9 &  33.5  & 1.3  & 3.4 \\
12 &  66.08 & 91.91 & 12.91 & 7.9   & 26.21 & 6.6 &  9.9 &  76.9  & 1.3  & 3.4 \\
14 &  69.92 & 77.64 & 10.29 & 7.0   & 21.27 & 4.2 &  8.1 &  30.6  & 1.1  & 2.8 \\
22 &  60.92 & 49.38 &  9.37 & 4.9   & 12.95 & 1.2 &  4.9 &  11.4  & 0.7  & 1.7 \\
27 &  51.18 & 69.67 & 11.56 & 4.2   & 10.91 & 2.0 &  4.1 &   9.4  & 0.6  & 1.4 \\
33 &  48.56 & 53.98 & 12.87 & 4.2   & 7.17  & 1.2 &  2.7 &   2.8  & 0.4  & 0.9 \\
41 &  48.10 & 65.58 & 11.61 & 3.8   & 6.74  & 1.6 &  2.6 &   2.9  & 0.3  & 0.9 \\
70 &  32.36 & 54.18 &  9.59 & 1.4   & 3.14  & 0.4 &  1.2 &   1.3  & 0.2  & 0.4 \\
73 &  30.28 & 44.96 & 11.51 & 1.5   & 2.56  & 0.3 &  1.0 &   1.0  & 0.1  & 0.3 \\
81 &  27.24 & 81.77 &  8.64 & 0.9   & 2.27  & 0.6 &  0.9 &   1.7  & 0.1  & 0.3 \\
84 &  28.80 & 47.84 & 11.82 & 1.4   & 2.40  & 0.3 &  0.9 &   0.8  & 0.1  & 0.3 \\ 
92 &  27.24 & 79.13 &  9.02 & 0.9   & 1.90  & 0.6 &  0.7 &   1.3  & 0.1  & 0.2 \\
103&  20.90 & 54.75 & 12.88 & 0.8   & 1.22  & 0.2 &  0.5 &   0.2  & 0.06 & 0.2 \\
\hline
\end{tabular}
\label{table_energy_1530}
\end{table*}

\subsection{Energetics of the shell}
The shell expansion velocities are taken as the mean values between 
the velocity separations of the red and blue components with respect to the central peak in each \hii\ region.
The values are shown in column~3 of Tables~\ref{table_energy_1530}-\ref{table_energy_6951}.
The adopted shell radius in parsecs ($\rm {R_{shell}=0.2~R_{reg}}$) for each \hii\ region is shown 
in column~2 of the same tables. Using the value of the shell electron density 
for each \hii\ region, we obtain the ionized mass of a thin shell integrating over its volume, thus:

\begin{equation} 
\rm M_{shell}(H^+)=4\pi R^{2}_{shell}\times\Delta R \times n_{shell}\times m_p
\end{equation}
where $\rm m_p$ is the mass of the hydrogen atom. The mass of the shell is shown in column~5, 
the values are of order $\rm 10^{4}\Msun$ and represent 18.9\%, 26.2\% and 24.2\% of the 
total ionized mass of the region for NGC~1530, NGC~3359 and NGC~6951 respectively. 
In column~7 is shown the kinetic energy of the shell, obtained using the formula
$\rm E_k=\frac{1}{2}m_{shell}v_{shell}^{2}$. The energies are of order $\sim 10^{51}$~erg, 
which is the same order as the kinetic energies of the most luminous shells in 
30~Doradus and NGC~604. 

We have compared the kinetic energy of the shell with input energies from the
central stars. We use the observed \ha\ luminosity of the region to derive an
equivalent number of O3(V) type stars (column 6 of Tables~\ref{table_energy_1530}-\ref{table_energy_6951}), 
using the \ha\ luminosity for an O3(V) type star given by Vacca et al. (1996). 
We will see that for models based on stellar winds as the principal source of
shell kinetic energy, the method gives only lower limit estimates of the input energy for two reasons: 
first, we do not take into account the photon escape fraction for the \hii\ regions, 
that can be as high as $\sim$80\% of the ionizing radiation for the most luminous \hii\ regions 
(see Zurita 2001), and second new recent effective temperatures for O(V) type 
stars would imply a reduction of the total logarithmic 
luminosity of these stars of $\sim$0.1 (Martins et al. 2002),  
larger equivalent number of O3(V) type stars obtained from the
observed \ha\ luminosity would imply higher energy inputs than those shown in 
Tables~\ref{table_energy_1530}-\ref{table_energy_6951}.   

Assuming a mean lifetime of $10^6$yr for the central ionizing stars, we can 
estimate their integrated wind energies, using the wind
luminosity for an O3(V) spectral type star, 
log~$\rm L_{wind}$= 37.08~(\ergs), given by Leitherer (1998). The values are 
shown in column~8 of Tables~\ref{table_energy_1530}-\ref{table_energy_6951}. 

For comparison with the wind energy input, we show in column~10 the energy of the 
ionizing radiation, obtained using the number of Lyman conti\-nuum 
ioni\-zing photons ($\rm 7.4\times10^{49}photon~s^{-1}$) 
for an O3(V) type star given by Vacca et al. (1996). Although recent studies 
by Martins et al. (2002) give somewhat lower values than these earlier studies, the 
overall energy budget for the ionizing radiation would in any case be 
higher than that of the winds by at least two orders of magnitude. In column~11 we  
show a first order approximation to the total radiative energy input for the \hii\ region assuming that 
the ionizing stars emit as blackbodies of temperature $\rm T_{eff}=51230~K$ and have a radius 
of $\rm R=13.2~\Rsun$, (Vacca et al. 1996).

In column~9 we show the turbulent kinetic energy of the \hii\ region. This is obtained 
from the total mass of the \hii\ region and the non--thermal velocity dispersion of the 
central and most intense component, $\rm E_{turb}=\frac{1}{2}M_{reg}\sigma_{nt}^{2}$. The 
values range from $10^{49}-10^{52}$~erg. There are 
some \hii\ regions whose central component is not completely resolved, which are marked as "==" 
in column~9 of Tables~\ref{table_energy_1530}-\ref{table_energy_6951}. These \hii\ regions 
show high intensity wing features located close to the central peak, whose lower expansion 
velocities could imply that the shells are located at greater distances from the 
centre of the region and that they could have been braked as they expand within the \hii\ region. Besides, 
the higher fractional Emission Measures of these features with respect to the fractional 
EM of other \hii\ regions means that they have more fractional mass in their shells. Thus, 
the corresponding fractional masses inside the region will be lower and thus the 
turbulent components and velocity dispersions should also be lower, as we find in practice.   

\subsection{Discussion}
The existence of an expanding shell in a significant fraction of \hii\ regions might appear 
to be in contradiction with the results on nearby extragalactic \hii\ regions where several shells 
with different radii have been observed. The contradiction is only apparent for two reasons: 

\begin{itemize}
\item[1.-] Both studies, for 30~Doradus and NGC~604, 
show a specific expanding shell which is much brighter than the other shells detected in these
regions and with similar radii in both cases, $\rm R_{shell}\sim 0.2~R_{reg}$, and expansion velocities
$\rm v_{exp}\sim70$~\kms\ for NGC~604 and $\rm v_{exp}\sim40$~\kms for 30~Doradus. The \ha\ luminosity of the 
prominent shell in NGC~604 represents $\sim$75\% of the total \ha\ luminosity of the observed expanding shells. 
Besides, the integrated line profiles for these \hii\ regions show extended wing features. In the case of 30~Doradus, 
Melnick et al. (1999) fitted the integrated line profile given by Chu \& Kennicutt (1994) with two 
Gaussian components, the more intense and narrower central one and another much broader. Their widths 
(after correcting from thermal and instrumental broade\-ning) are $\rm\sigma=22$~\kms and 
$\rm\sigma=44$~\kms\ for the narrow and broad one respectively. Melnick et al. (1999) interpreted this
broad feature as a single turbulent motion of unknown origin. We interpret here   
this broadest Gaussian component as a shell with an expansion velocity of 
$\rm v_{exp}\sim45$~\kms\ moving inside the \hii\ region as shown in a spectrum with either too low a velocity
resolution or too low a S:N ratio to resolve the wing features. In fact, the line profile extracted by 
Smith \& Weedman (1972) from Fabry-P\'erot observations in the center of the region shows line profiles with two 
distinct peaks separated by 45~\kms. 
\item[2.-] In the case of 30~Doradus the slit--spectra observations are limited to cover a 
small fraction of the region. As Chu \& Kennicutt (1994) pointed out, the background velocity field 
they observed in the slit--spectra did not allow them to detect any expanding shell with 
$\rm v_{exp}\leq 50$~\kms, unless it is the dominant emission feature along the line of sight. In our 
observations we do not have this limitation. If we are observing an expanding shell that does 
not have a high surface brightness but does occupy a significant fraction of the
projected area of the \hii\ region, the best way to distinguish it clearly from the emission of 
the rest of the \hii\ region is to obtain two--dimensional spectra which can integrate the emission 
of the shell over a significant fraction of its area and hence can separate it in velocity space from the 
emi\-ssion of the rest of the region. This is the process we have followed here to
detect the expanding shells in our regions. 
\end{itemize}
Based on the two points explained above we believe that the high velocity features 
observed in the line profiles of our sample of \hii\ regions are produced by the emission 
of expanding shells with similar characteristics to those with high luminosities 
observed in 30~Doradus and NGC~604. 
Although we do not reject the possibility that there may be other shells inside the \hii\ regions, 
as they have been indeed detected for these nearby regions, we believe that our collective 
evidence demonstrates that these do not contribute significantly 
to the emission of the high velocity features observed in the integrated line profiles of the \hii\ 
regions over their full surfaces.

%\clearpage

\begin{table*}[]
\centering
%\hspace{-4.0cm}
%\small
\caption[]{The same parameters as in Table~\ref{table_energy_1530} but for \hii\
  regions NGC 3359.}
%\vspace{0.5cm} 
%\hspace{-2.0cm}
\begin{tabular}{ccccccccccc}
\hline
\hline
\scriptsize Region & \scriptsize $\rm R_{shell}$ & \scriptsize $\rm v_{shell}$ & \scriptsize $\rm n_{shell}$ 
& \scriptsize $\rm M_{shell}$ & \scriptsize $\rm N_{eq}$ & \scriptsize $\rm E_K$ & \scriptsize $\rm E_{wind}(O3)$ 
&\scriptsize $\rm E_{turb}$ &\scriptsize  $\rm E_{rad}$ & \scriptsize $\rm Etotal_{rad}$ \\
\scriptsize (number) & \scriptsize (pc) & \scriptsize (\kms) &\scriptsize (\cmtres) & \scriptsize ($\rm 10^{4}\Msun$) 
& \scriptsize (O3(V)) &
\scriptsize ($10^{51}$~erg) & \scriptsize ($10^{51}$~erg) & \scriptsize ($10^{51}$~erg) & 
\scriptsize ($10^{54}$~erg) & \scriptsize ($10^{54}$~erg) \\
\hline
1   & 97.70 & 63.63 & 10.02 & 13.4& 36.81 & 5.4  & 14.0  & 16.8 & 1.9  & 4.8 \\
6   & 69.62 & 63.97 &  8.36 & 56.6& 13.97 & 2.3  &  5.3  & 7.1  & 0.7  & 1.8 \\
15  & 68.00 & 81.91 &  3.63 & 2.3 & 9.42  & 1.6  &  3.6  & 7.6  & 0.5  & 1.2 \\
16  & 68.40 & 84.56 &  3.23 & 2.1 & 8.55  & 1.5  &  3.2  & 6.4  & 0.4  & 1.1 \\
17  & 52.00 & 67.31 &  8.04 & 3.0 & 8.11  & 1.4  &  3.1  & 5.8  & 0.4  & 1.1 \\
19  & 63.30 & 41.85 & 10.39 & 5.8 & 8.03  & 1.0  &  3.0  & ===  & 0.4  & 1.0 \\
38  & 47.70 & 45.95 &  7.83 & 2.5 & 4.21  & 0.5  &  1.6  & 0.4  & 0.2  & 0.6 \\
42  & 48.66 & 40.91 &  8.16 & 2.7 & 3.71  & 0.5  &  1.4  & ===  & 0.2  & 0.5 \\ 
44  & 52.34 & 59.17 &  3.88 & 1.5 & 3.66  & 0.5  &  1.4  & 0.8  & 0.2  & 0.5 \\ 
45  & 51.08 & 43.81 &  7.76 & 2.8 & 3.50  & 0.8  &  1.3  & ===  & 0.2  & 0.5 \\
47  & 45.28 & 77.29 &  4.63 & 1.3 & 3.42  & 0.4  &  1.3  & 2.0  & 0.2  & 0.4 \\
51  & 37.14 & 46.73 &  9.42 & 1.8 & 3.23  & 0.3  &  1.2  & 0.05 & 0.2  & 0.4 \\
52  & 34.54 & 44.97 &  8.58 & 1.4 & 3.18  & 0.5  &  1.2  & ===  & 0.2  & 0.4 \\
53  & 54.62 & 61.05 &  3.27 & 1.3 & 3.04  & 0.5  &  1.2  & 0.3  & 0.2  & 0.4 \\
56  & 51.64 & 44.48 &  6.44 & 2.4 & 2.82  & 0.5  &  1.1  & ===  & 0.1  & 0.4 \\
83  & 39.10 & 41.19 &  7.29 & 1.6 & 1.76  & 0.3  &  0.7  & ===  & 0.09 & 0.2 \\
92  & 36.38 & 51.98 &  4.80 & 0.9 & 1.51  & 0.2  &  0.6  & ===  & 0.08 & 0.2 \\
103 & 37.14 & 55.86 &  4.37 & 0.8 & 1.28  & 0.3  &  0.5  & 0.7  & 0.07 & 0.2 \\
140 & 34.54 & 51.17 &  6.27 & 1.0 & 0.85  & 0.3  &  0.3  & 0.1  & 0.04 & 0.1 \\
155 & 30.84 & 49.61 &  4.05 & 0.5 & 0.75  & 0.1  &  0.3  & ===  & 0.04 & 0.01 \\
192 & 26.96 & 55.15 &  3.86 & 0.4 & 0.53  & 0.1  &  0.2  & 0.05 & 0.03 & 0.07 \\
\hline
\end{tabular}
\label{table_energy_3359}
\end{table*}

\begin{table*}[]
\centering
%\hspace{-3.0cm}
%\small
\caption[]{The same parameters as in Table~\ref{table_energy_1530} and 
Table~\ref{table_energy_3359} but for \hii\
  regions NGC 6951.}
\begin{tabular}{ccccccccccc}
\hline
\hline
\scriptsize Region & \scriptsize $\rm R_{shell}$ & \scriptsize $\rm v_{shell}$ & \scriptsize $\rm n_{shell}$ 
& \scriptsize $\rm M_{shell}$ & \scriptsize $\rm N_{eq}$ & \scriptsize $\rm E_K$ & \scriptsize $\rm E_{wind}(O3)$ 
&\scriptsize $\rm E_{turb}$ &\scriptsize  $\rm E_{rad}$ & \scriptsize $\rm Etotal_{rad}$ \\
\scriptsize (number) & \scriptsize (pc) & \scriptsize (\kms) &\scriptsize (\cmtres) & \scriptsize ($\rm 10^{4}\Msun$) 
& \scriptsize (O3(V)) &
\scriptsize ($10^{51}$~erg) & \scriptsize ($10^{51}$~erg) & \scriptsize ($10^{51}$~erg) & 
\scriptsize ($10^{54}$~erg) & \scriptsize ($10^{54}$~erg) \\
\hline
2  & 66.62  & 60.93 & 11.11 & 6.9 &18.11 & 2.5 & 6.9  & 7.1 & 0.9  & 2.4 \\
4  & 54.80  & 52.08 & 11.17 & 4.7 &12.61 & 1.3 & 4.8  & 3.8 & 0.6  & 1.6 \\
6  & 54.92  & 47.09 & 11.02 & 4.6 & 9.17 & 1.0 & 3.5  & 2.5 & 0.5  & 1.2 \\
10 & 51.50  & 41.73 &  9.75 & 3.6 & 6.74 & 0.6 & 2.6  & 0.2 & 0.3  & 0.9 \\
11 & 52.14  & 60.47 &  8.53 & 3.2 & 6.48 & 1.2 & 2.5  & 1.8 & 0.3  & 0.8 \\ 
12 & 52.02  & 65.41 &  8.07 & 3.1 & 6.27 & 1.3 & 2.4  & 3.5 & 0.3  & 0.8 \\ 
13 & 52.02  & 46.71 &  8.53 & 3.2 & 6.26 & 0.7 & 2.4  & 2.4 & 0.3  & 0.8 \\
15 & 49.76  & 76.36 &  5.44 & 1.9 & 5.38 & 1.1 & 2.0  & 4.2 & 0.3  & 0.7 \\
18 & 48.66  & 47.62 &  6.50 & 2.2 & 4.58 & 0.5 & 1.7  & 0.6 & 0.2  & 0.6 \\ 
24 & 37.14  & 50.95 &  8.23 & 1.6 & 3.29 & 0.4 & 1.2  & 0.4 & 0.2  & 0.4 \\ 
27 & 40.12  & 48.45 &  7.31 & 1.6 & 2.79 & 0.4 & 1.1  & 0.2 & 0.1  & 0.4 \\ 
36 & 36.78  & 38.62 &  7.97 & 1.5 & 2.31 & 0.2 & 0.9  & === & 0.1  & 0.3 \\ 
41 & 37.34  & 50.82 &  8.27 & 1.6 & 2.20 & 0.4 & 0.8  & 0.4 & 0.1  & 0.3 \\
62 & 31.22  & 84.12 &  6.34 & 0.9 & 1.45 & 0.6 & 0.5  & 0.8 & 0.07 & 0.2 \\
67 & 27.52  & 42.83 &  8.17 & 0.9 & 1.32 & 0.2 & 0.5  & === & 0.07 & 0.2 \\
69 & 31.00  & 78.53 &  5.79 & 0.7 & 1.26 & 0.5 & 0.5  & 1.4 & 0.07 & 0.2 \\
75 & 32.28  & 51.53 &  6.45 & 0.9 & 1.13 & 0.2 & 0.4  & 0.2 & 0.06 & 0.1 \\
\hline
\end{tabular}
\label{table_energy_6951}
\end{table*}

\section{Feasible driving mechanisms}
\subsection{Stellar winds from the ionizing stars}
The winds coming from the ionizing stars interact with the ISM that su\-rrounds the stars. 
The interaction produces a thin shell of swept--up interestellar medium with 
less dense material in its interior, which is referred to as a bubble.

The dynamics of this types of bubbles was modelled in considerable detail by
Dyson (1980), who found that the kinetic energy of the shell is related to the stellar wind luminosity 
($\rm\dot E_{w}={1\over 2} \dot M_{w}v_{w}^{2}$), $\rm E_{k}=0.20\dot E_{w}t$, where t 
is the lifetime for the central ionizing star. Thus, we would expect a ratio between the 
kinetic energy of the shell and the kinetic energy of the wind of $\sim~$0.2.

Assuming a mean lifetime of $10^6$yr for the central ionizing stars, we can 
estimate their integrated wind energies, using the wind
luminosity for an O3(V) spectral type star, 
log~$\rm L_{wind}$= 37.08~(\ergs), given by Leitherer (1998). The values are 
shown in column~8 of Tables~\ref{table_energy_1530}-\ref{table_energy_6951}. 

\begin{figure*}
\centering
\includegraphics[width=10cm]{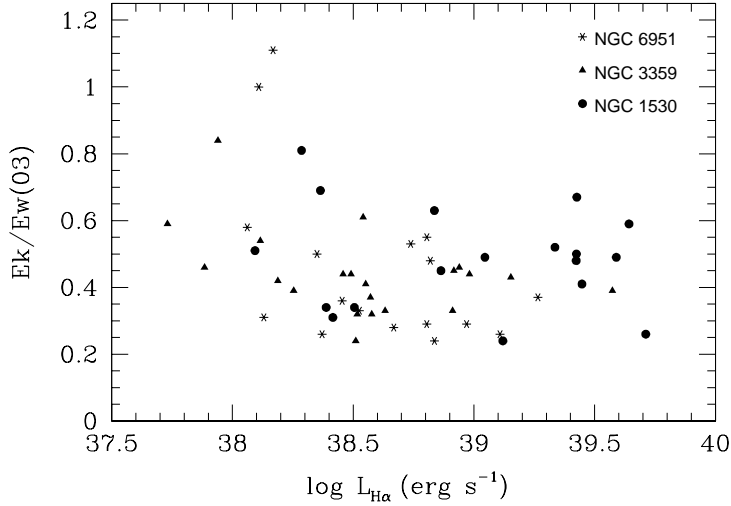}
\protect\caption[ ]{Ratio between the shell kinetic energy and the 
stellar input wind energy {\it versus} logarithmic \ha\ luminosity for 
the \hii\ region sample with high velocity features in NGC~1530, NGC~3359 and NGC~6951.}
\label{Ekvlum}
\end{figure*}

\begin{figure*}
\centering
\includegraphics[width=10cm]{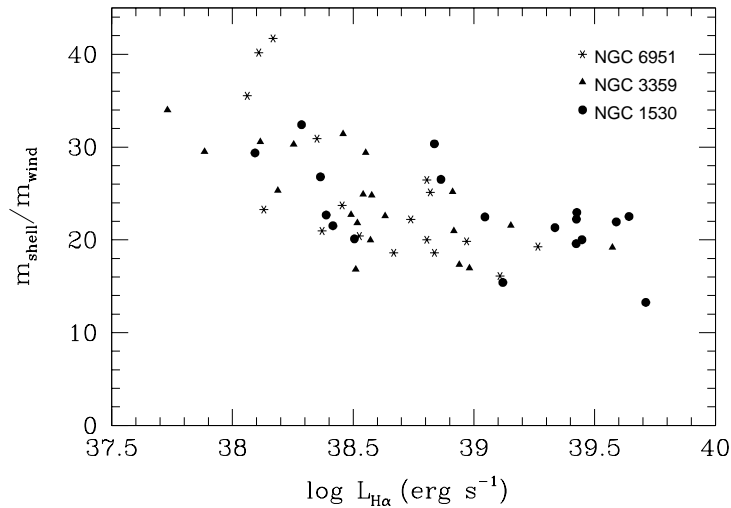}
\protect\caption[ ]{Ratio between the shell momentum and the wind 
momentum {\it versus} logarithmic \ha\ luminosity.}
\label{momvlum}
\end{figure*}

In Fig.~\ref{Ekvlum} we show the ratio between the shell 
kinetic ener\-gy and the stellar input wind energy {\it versus} logarithmic \ha\ luminosity for 
the \hii\ region sample. As can be seen in the figure, 
most of the ratios are in the range 0.25-0.6. 
The momentum of the shell can also be estimated with our observations, using the calculated shell
 mass and expansion velocity, 
$\rm m_{s}=M_{shell}\times v_{shell}$. The wind momentum can be obtained 
with the equivalent number of O3(V) type stars given in column~6 of 
Tables~\ref{table_energy_1530}-\ref{table_energy_6951} and the value 
given by Leitherer (1998) for an O3(V) type star, $\rm m_{w}=7.56\times 10^{28}~g~cm~s^{-1}$.   
In Fig.~\ref{momvlum}, we show the ratio $\rm m_{s} \over m_{w}$, typical values ranges from 15 to 30.

The values of $\rm E_{\rm k}\over E_{\rm wind}$ are somewhat higher than the theoretical fraction 
obtained above. The differences with respect to the theoretical ratio 0.2 could well be due to the uncertainties 
in obtaining a reliable value for the wind energy input, based on the fact that we are obtaining it from the 
observed \ha\ lu\-mi\-no\-si\-ty and that a significant fraction of the ionizing luminosity 
is probably leaking. The observed ratios $\rm m_{s} \over m_{w}$ are also affected by the difficulties 
in obtaining a reliable value of the stellar wind momentum using the equivalent number of O3(V) type stars. 
Using a typical value of the velocity of the shell of 80~\kms\  and 
$\rm v_{wind}=3200~\kms$ (Leitherer 1998) for the wind velocity of an O3(V) type star, the 
expected ratio of $\rm m_{s} \over m_{w}$ for the theoretical ratio 
$\rm E_{\rm k}\over E_{\rm wind}$~$=0.2$ shown 
above, is $\rm m_{s} \over m_{w}$~$\sim 8$. In any case, the ener\-gy 
and momentum ratios observed show that the stellar winds could 
be responsible for driving the shell out to radii of order of $\rm 0.2~R_{\rm reg}$ 
and that the differences might rea\-so\-na\-bly be due to the uncertainties 
in obtaining a reliable value for the wind energy input, as well as the
uncertainties entailed in deriving the shell energies and momenta from our observations.

\subsection{Shells with larger radii}
We have investigated further the possibility that the shells might 
have larger radii than those assumed from nearby \hii\ regions.
For two \hii\ regions of each galaxy, we have obtained
profiles of $1\times1$~pixels of aperture, located at $\rm\sim 0.5~R_{reg}$. 
The \hii\ regions which permit us to
obtain these profiles with a significant S:N are among the biggest and most luminous \hii\ regions in the
sample. In all of the line profiles extracted as examples, we have found wing features 
with similar velocities as those found for the best S:N line profiles. This means that for these high luminosity
\hii\ regions the expanding shell could be at least at a radius of $\rm\sim 0.5~R_{reg}$.

For these \hii\ regions we have computed the shell electron densities and energy requirements 
assuming that the shells are located at $\rm R_{s}=0.5~R_{reg}$. The results are shown in 
Table~\ref{table_reg0.5}. All of the ratios, between the kinetic energy of the shell and the wind energy input, 
$\rm E_K/\rm E_{wind}(O3)$, shown in column~6 of this table have ratios higher
than 0.6, in fact the highest ratio is 1.48. 
These high values are not easily explained with the stellar wind mechanism. 
Thus, there are probably other mechanisms that can transfer energy to the ISM to
produce the kinetic energy of these observed shells.

\begin{table}[]
\hspace{-3.0cm}
%\small
\caption[]{Energetics of the shell for two high luminosity \hii\ regions in each galaxy with a shell radius of
$\rm0.5~R_{reg}$. Column 1: \hii\ region number from the catalogue, in parenthesis we show the galaxy they
belong to. Column 2: Logarithmic \ha\ luminosity. Column 3: Shell radius ($\rm0.5~R_{reg}$).
Column 4: Shell expansion velocity. Column 5: Mean shell electron density. 
Column 6: Ratio between the kinetic energy of the shell and the wind energy input 
from the equivalent O3(V) spectral type stars obtained to produce the observed \ha\ luminosity.}
%\vspace{0.5cm} 
%\hspace{-0.5cm}
%\centering
\begin{tabular}{cccccc}
\hline
\hline
\scriptsize Region & \scriptsize  log~L$_{\scriptsize\ha}$ & \scriptsize $\rm R_{shell}$ 
& \scriptsize $\rm v_{shell}$ & \scriptsize $\rm N_{shell}$ 
& \scriptsize $\rm E_K/\rm E_{wind}(O3)$ \\
\scriptsize (number) & \scriptsize (\ergs) &\scriptsize (pc) & \scriptsize (\kms) &\scriptsize (\cmtres) & \\
\hline
6 (N1530)	& 39.64 & 219.15 & 83.10 & 5.23 & 1.48 \\
27(N1530)	& 39.04 & 127.95 & 69.67 & 4.63 & 1.23 \\
1 (N3359)	& 39.57 & 244.25 & 63.63 & 4.01 & 0.96 \\
19(N3359)	& 39.91 & 158.25 & 41.85 & 4.15 & 0.83 \\ 
2(N6951)	& 39.27 & 166.50 & 60.93 & 4.44 & 0.93 \\
10(N6951)	& 38.84 & 128.75 & 41.73 & 3.90 & 0.61 \\
\hline
\end{tabular}
\label{table_reg0.5}
\end{table}

We suggest two mechanisms which are possible in the context of the known stellar 
properties of the ionizing clusters, which 
are supernova energy injection and radiation dri\-ving via dust coupling to the 
gas following the formalism applied by Elitzur \& Ivezic (2001). 
Supernovae release large amounts of energy, typically of $\rm E\sim10^{51}$~erg, but the efficiency of 
conversion of this energy into kinetic energy in the ISM is $\sim 4$\% (Dyson
1980), which is too low to offer a canonical
explanation for the shell kinetic energies measured. 
The stellar radiation from the ionizing stars offers a big energy source which
might be available to drive the shell. The stellar radiation from the ionizing 
stars represents a major energy source
which would require only a low coupling efficiency to provide a valid additional
input of outflow momentum to the shell. Dust coupling might be one possibility, but
the absorption coefficient is known to be low. The photoionization process itself 
can also contribute. Although the coupling efficiency here is also quite 
low (Dyson 1980), and this source alone would not drive out material at more 
than a few times the sound speed, it might well provide an
effective incremental contribution which would enhance the effect of the
stellar winds (which we have discussed more quantitatively above), enabling the flow to
persist out to larger radii with higher overall outflow momentum than that provided
by the winds alone. Further consideration of these mechanisms calls for a detailed
modelling effort which is beyond the scope of the present paper. 

As can be seen from comparison of column~7 and columns~10 and~11 in 
Tables~\ref{table_energy_1530}-\ref{table_energy_6951}, 
the total radiative energy emitted by the central stars is some three orders of
magnitude higher than the kinetic energy of a
characteristic expanding shell. In any case, this energy must be taken 
into account when trying to find an explanation 
for the high kinetic energies of the shells observed at relatively large radii inside
the \hii\ regions. However, in this essentially observational study we can only point
the possible mechanisms which might explain the expanding shells 
located at large radii inside the \hii\ region, for which the stellar winds may not provide 
energy enough to drive the shell. 

\section{Conclusions}
\begin{itemize}
\item We have studied the existence of low intensity high ve\-lo\-ci\-ty components 
in the integrated line 
profiles of the \hii\ region populations of three barred spiral galaxies. We have analyzed the 
integrated line profiles and extracted a representative sample of 
spectra which show these components. Down to a limit of log~L$_{\scriptsize\ha}=38.0$~(\ergs), we 
found that 26.77\%, 42.0\% and 31.63\% of the catalogued \hii\ regions in 
NGC~3359, NGC~6951 and NGC~1530, respectively, show high velocity low intensity features in their 
integrated line profiles. 

\item The higher fraction is correlated with a higher S:N ratio in the 
line profiles, which suggests that at sufficiently high value of S:N all \hii\ 
regions may well show these wing features, which means that these high velocity
features could well exist in most \hii\ region line profiles, and as such
provide evi\-den\-ce of a widely distributed phenomenon. 

\item Based on the mean S:N of the two high velocity components, we have obtained
  the spectra from the set of line profiles
with increasing apertures that best defines the wing features and used them 
as the representative spectra of each region which characterize the wing parameters.  

\item The analysis of the fit parameters of the high velocity low intensity features 
show that parameters of the red and blue component are in the same range of values. 
The mean va\-lues of the Emission Measure for both components are 8.9\% of the 
total EM of the \hii\ region and the velocity separation ranges from 40~\kms\ to 90~\kms\ 
for our sample of the most luminous \hii\ regions (which cover the high of the
luminosity range), with a mean value for each
component of $<\rm v_{red}>=60.5$~\kms\ and $<\rm v_{blue}>=59.02$~\kms. The
mean measured velocity dispersions 
for the red and blue components are $<\rm\sigma_{red}>=19.5$~\kms\ and
$<\rm \sigma_{blue}>=17.0$~\kms, though some line profiles are not fully
resolved.  

\item We interpret the high velocity features as evidence of an ex\-pan\-ding
  shell inside the \hii\ region. The
shell expansion velocity is obtained as the mean value of $\rm v_{red}$ and $\rm v_{blue}$ for each \hii\ region. 
Based on parameters for the highest luminosity shells observed in the nearby extragalactic \hii\ regions, 
30~Doradus and NGC~604, we have estimated the physical parameters and the energetics of the shell of 
each \hii\ region in our sample. 

\item We have compared the shell kinetic energy with the kinetic ener\-gy of the 
stellar winds coming from the ionizing stars inside the \hii\ region. For shells with radii 
$\rm\sim 0.2~R_{reg}$, the ratios we derive between the kinetic energy of the shell and the energies of the 
stellar winds are rather higher than the expected theoretical ratio 0.2, 
but the uncertainties in obtaining a reliable
value for the wind energy input from the observed \ha\ luminosity, as well as
those inherent in our derivation of the observational values, can well explain the differences. 
Thus, the ratios obtained imply that ste\-llar winds 
alone could drive the expanding shells, if these are located within $\rm\sim 0.2~R_{reg}$. 

\item Some of the most luminous \hii\ regions of our sample show evidence that their expanding shells  
are located at radii higher than $\rm\sim 0.5~R_{reg}$. For these \hii\ regions
the stellar wind mechanism alone may well not be able to explain 
the observed shell kinetic energies or momenta, implying that there are other mechanisms
required to account for the observed properties.

\end{itemize}

%__________________________________________________________________

\begin{acknowledgements}
This work was supported by the Spanish DGES (Direcci\'on General de 
Ense\~nanza Superior) via Grants PB91-0525, PB94-1107 and PB97-0219 
and by the Ministry of Science and Technology via grant AYA2001-0435 and AYA2004-08251-C02-01.  
The WHT is o\-pe\-ra\-ted on the island of La Palma by the Isaac Newton Group in
the Spanish Observatorio del Roque de los Muchachos of the Instituto de Astrof\'\i sica de 
Canarias. We thank the referee, Dr. C.R.O'Dell, for his constructive comments which helped us 
 to make significant improvements to the article.
\end{acknowledgements}

\end{document}